\documentclass[12pt]{article}
\usepackage{graphicx}
\usepackage{enumerate}
\usepackage[agsm,abbr]{harvard}
\newcommand{\boldomega}{\mbox{\boldmath{$\omega$}}}

\newcommand{\boldrho}{\mbox{\boldmath{$\rho$}}}
\newcommand{\boldphi}{\mbox{\boldmath{$\phi$}}}

\newcommand{\bu}{\mathbf{u}}
\newcommand{\bw}{\mathbf{w}}

\newcommand{\bz}{\mathbf{z}}

\title{Block clustering with collapsed latent block models}
\author{Jason Wyse and Nial Friel \\ University College Dublin, Belfield, Dublin 4, Ireland }

\begin{document}

\maketitle

\begin{abstract}
We introduce a Bayesian extension of the latent block model for model-based block clustering of data matrices. Our approach considers a block model where block parameters may be integrated out. The result is a posterior defined over the number of clusters in rows and columns and cluster memberships. The number of row and column clusters need not be known in advance as these are sampled along with cluster memberhips using Markov chain Monte Carlo. This differs from existing work on latent block models, where the number of clusters is assumed known or is chosen using some information criteria. We analyze both simulated and real data to validate the technique. 
\end{abstract}

\noindent {\bf Keywords:} Block clustering, latent block model, Bayesian model choice, collapsed model

\section{Introduction}

Many data sets arise as a result of a number of features or variables being observed for a collection of objects. As examples; shoppers and the items which they do or do not buy; whether a document contains specific words or not; the expression levels of a gene under a series of conditions in a DNA experiment. Such data will be recorded in a matrix, say, with rows indexing objects and columns indexing features or variables. Often interest will focus on clustering rows and further, clustering the features which distinguish these row clusters. We refer to this task as block clustering although it is also known as block modelling, biclustering, co-clustering and two-mode clustering.

One of the first approaches to block clustering was suggested by~\citeasnoun{Hartigan72} and since then, many have been proposed. Much recent work in block clustering and related areas has been either on the analysis of microarray data~\cite{Tibshirani99,Cheng00,Getz00,Lazzeroni02,Kluger03} or document classification \cite{Hofmann01,Blei03,Griffiths04}. Approaches vary in whether they allow clusters to overlap or not. In our case, the problem can be thought of as permuting the rows and columns of the data matrix to make a ``chessboard'' of blocks of data having similar value.

While many approaches to document classification are model-based $\mbox{i.e.}$ a parametric underlying model is assumed when clustering data, this is often not the case in microarray analysis. Some exceptions are~\citeasnoun{Lazzeroni02} who assume a Gaussian error model for gene expression with additive effects for gene and condition clusters and~\citeasnoun{Sheng03} who assume a multinomial model for expression level in a discretized microarray. It is common to use two-way hierarchical clustering for this data or other partitioning methods (for example~\citeasnoun{Getz00}). One drawback of these methods is the lack of probabilistic justification as noted by~\citeasnoun{Wit04} (Chapter 7, page 171). A model-based approach allows explicit modelling of noise in the data. This can be an advantage in data, such as microarrays, which is particularly prone to noise, incorporating uncertainty in cluster membership.

In this paper we consider an extension of the latent block model (LBM) approach of~\citeasnoun{Govaert08}. The LBM was developed as an intuitive extension of the finite mixture model used in model-based clustering~\cite{Fraley02} to allow clustering of objects and features.  We propose a Bayesian LBM. This has been considered previously by~\citeasnoun{vanDijk09}. In their approach the number of clusters in objects and features is assumed known and Gibbs sampling is used to find clusterings. They choose the number of clusters using an information criterion based on maximum likelihood. 

We show that it is possible to sample the number of clusters and the cluster membership jointly using simple Markov chain Monte Carlo (MCMC) on a collapsed model, so that uncertainty in the number of clusters is naturally incorporated as part of our Bayesian LBM. The collapsed model is obtained by integrating out block parameters analytically. This is possible using standard prior assumptions. There is no need to resort to a trans-dimensional sampler, such as the reversible jump sampler of~\citeasnoun{Green95}. Our idea extends the allocation sampler of~\citeasnoun{Nobile07} to two directions, with slight modifications. We discuss differences and similarities of our approach to block clustering with those which are most comparable qualitatively. The sampler is applied to both simulated and real datasets to gauge performance.

The remainder of the paper is organized as follows. Section 2 reviews the LBM and introduces the collapsed Bayesian LBM. Section 3 gives the MCMC sampler which we use. The differences between our sampler and reversible jump samplers are also discussed. Label switching is mentioned and the section concludes with approaches to summarize the output of the sampler. Section 4 applies the approach to simulated data. In Section 5 we analyze voting records data from the U.S. congress and compare this to a maximum likelihood analysis. In Section 6 we discuss the analysis of microarray data, and use our sampler to analyze data from a yeast microarray experiment. The paper concludes with a discussion. 

\section{Models}

The data is $Y=(y_{ij})$, an $n\times m$ matrix. It is assumed rows and columns may be reordered so that the matrix can be represented as $K\times G$ blocks with data in blocks modelled by the same density, where $K$ and $G$ are the number of row and column clusters respectively. This could be imagined as a ``chessboard'' effect, with $K-1$ divisions in the direction of the rows and $G-1$ in the direction of the columns. The parameters of the data density are conditional on the block and $\theta_{kg}$ denotes the parameters for block $(k,g)$, with $\Theta$ denoting the collection of these. We now give a review of the LBM of~\citeasnoun{Govaert08}.

\subsection{Latent block models}

Conditional on $K$ and $G$, let $\mathcal{U}$ be a latent space indexing the set of all possible clusterings of rows and columns. Then the distribution of the data $Y$ can be written
\[
p(Y|K,G,\Theta,\boldphi) = \sum_{\bu\in \mathcal{U}} p(\bu|K,G,\boldphi) p(Y|K,G,\bu,\Theta)
\]
where $\boldphi$ are parameters for the distribution of $\bu$.~\citeasnoun{Govaert08} make the assumption that row and column clusterings are independent {\it a priori}, so that $p(\bu|K,G,\boldphi)=p(\bz|K,\boldomega)p(\bw|G,\boldrho)$ where $z_i=k$ if row $i$ is in cluster $k$ and $w_j=g$ if column $j$ is in cluster $g$. The probability of a row belonging to cluster $k$ is $\omega_k$ and $\rho_g$ denotes the probability that a column belongs to cluster $g$. The LBM is then
\begin{eqnarray}
p(Y|K,G,\Theta,\boldomega,\boldrho) & = & \sum_{(\bz,\bw)\in\mathcal{Z}\times\mathcal{W}} p(\bz|K,\boldomega)p(\bw|G,\boldrho) p(Y|K,G,\bz,\bw,\Theta) \label{eq:data_like}
\end{eqnarray}
where $\mathcal{Z}$ and $\mathcal{W}$ denote the latent spaces of all row and column clusterings respectively. 
When constructing the data likelihood given the latent allocations, we make the assumption of local independence. That is, within a block, data are independent. This gives data likelihood conditional on $\bz,\bw$,
\[
p(Y|K,G,\bz,\bw,\Theta) = \prod_{k=1}^K \prod_{g=1}^G \prod_{i:z_i=k} \prod_{j:w_j=g} p(y_{ij}|\theta_{kg}).
\]

As $|\mathcal{Z}\times\mathcal{W}|=K^{n}G^{m}$, it is not feasible to calculate (\ref{eq:data_like}). We now review a way to fit this model using a method based on Expectation-Maximization (EM)~\cite{Dempster77} due to~\citeasnoun{Govaert08}.

\subsubsection{Estimation using BEM2} \label{sec:EM_explanation}

\noindent Here we outline the BEM2 algorithm of~\citeasnoun{Govaert08} which we will compare our approach with later (Section~\ref{sec:congressional_voting}). In the interests of avoiding ambiguities in notation, we introduce the matrices $\mathbf{r} = (r_{ik})$ and $\mathbf{c} = (c_{jg})$ such that if row $i$ is in cluster $k$, $r_{ik}=1$ and 0 otherwise. Similarly for $c_{jg}$.  The complete (or classification) log-likelihood associated with the LBM (\ref{eq:data_like}) is 
\begin{equation}
\mathcal{L}(\mathbf{r},\mathbf{c},\boldomega,\boldrho,\Theta) = \sum_{i=1}^n \sum_{k=1}^K r_{ik} \log \omega_k + \sum_{j=1}^m \sum_{g=1}^G c_{jg} \log \rho_g + \sum_{i=1}^n \sum_{j=1}^m \sum_{k=1}^K \sum_{g=1}^G r_{ik} c_{jg} \log p(y_{ij}|\theta_{kg}).
\end{equation}
The E step using this log-likelihood directly is intractable due to the dependence structure among the rows and columns.~\citeasnoun{Govaert08} suggest a variational approximation to the joint distribution of the latent $\mathbf{r},\mathbf{c}$ which leaves $\mathbf{r}$ and $\mathbf{c}$ independent. Then using the interpretation of EM due to~\citeasnoun{Neal98} this leads to a new ``fuzzy'' criterion for block clustering
\[
\mathcal{G}(\mathbf{s},\mathbf{t},\boldomega,\boldrho,\Theta) = \mathcal{L}(\mathbf{s},\mathbf{t},\boldomega,\boldrho,\Theta) - \sum_{i=1}^n \sum_{k=1}^K s_{ik}\log s_{ik} - \sum_{j=1}^m \sum_{g=1}^G t_{jg} \log t_{jg}
\]
which can be alternately maximized with respect to $\mathbf{s}, \mathbf{t}$ and $\boldomega,\boldrho,\Theta$ where $s_{ik} = \Pr(z_i=k)$ and $t_{jg}=\Pr(w_j = g)$.

The possible ways in which this criterion may be alternately maximized determines different algorithms. The BEM2 algorithm maximizes it as follows.
\begin{enumerate}
\item Initialize the unknowns $\mathbf{s},\mathbf{t},\boldomega,\boldrho,\Theta$ at some sensible value.
\item Maximize $\mathcal{G}$ with respect to $\mathbf{s},\boldomega$ and $\Theta$ keeping $\mathbf{t}$ and $\boldrho$ fixed.
\item Maximize $\mathcal{G}$ with respect to $\mathbf{t},\boldrho$ and $\Theta$ keeping $\mathbf{s}$ and $\boldomega$ fixed.
\item Iterate steps 2-3 until convergence.
\end{enumerate}
It is noted that each sweep of BEM2 has two maximizations of $\Theta$. This maximization procedure is reported to have outperformed the other schemes considered in~\citeasnoun{Govaert08}, so we use it here to compare with our Bayesian approach.

\subsubsection{Choosing $K$ and $G$ when using BEM2} \label{sec:determine_k_and_g}

In (\ref{eq:data_like}), it is assumed that $K$ and $G$ are known. The number of clusters assumed can have a considerable effect on the output of clustering algorithms. Usually, many runs, each with a different number of clusters, are necessary. These are then compared to find the best clustering, either based on some information criterion or visual inspection of plots.

Since the LBM is defined in terms of the latent allocation vectors $\bz$ and $\bw$, it is not clear how one could use a standard information criterion ($\mbox{e.g.}$ BIC~\cite{Schwarz78}) here to choose the number of components best supported by the data. Evaluation of the log density of $Y$ at the MLE requires a sum over all $K^n G^m$ terms as in (\ref{eq:data_like}), which even for the moderate case of $K=2,G=2,n=10,m=10$ would require the sum of roughly one million log-likelihood evaluations. An alternative may be to use the maximized complete log-likelihood treating the row and column allocations as unknown parameters, $\mathcal{L}(\hat{\mathbf{r}},\hat{\mathbf{c}},\hat{\boldomega},\hat{\boldrho},\hat{\Theta})$ .~\citeasnoun{vanDijk09} have used this approach for LBMs when using AIC-3~\cite{Bozdogan94} to choose $K$ and $G$. In this case the number of parameters to be estimated is $n(K-1)+m(G-1)+dKG+(K-1)+(G-1)$ where $d$ is the dimension of any $\theta_{kg}$. A separate model estimation is required for each $K$ and $G$ combination over a grid of plausible models. Adopting this type of approach  crucially involves replacing the maximum log likelihood with a maximized complete log likelihood and also the issue of selection of a particular information criterion, and could therefore be criticized for both reasons.

The Bayesian LBM we propose seeks to incorporate uncertainty in $K$ and $G$ into the model. This is so that the clustering task is also one of cluster model determination. The model determination task and the allocation task are dealt with simultaneously through a fully Bayesian approach. This has analogy with some other block clustering strategies, which undertake greedy searches to find new row and column clusters. See for example~\citeasnoun{Hartigan00}. An advantage here is that the search has a probabilistic justification based on a posterior distribution for $K$ and $G$. In the next section we introduce the Bayesian LBM which is at the core of the clustering procedures we discuss.

\subsection{Bayesian latent block models} \label{sec:BayesLBM}

The Bayesian LBM is formed by taking prior densities on $K$, $G$, $\Theta$, $\boldomega$ and $\boldrho$. Let $\pi(\cdot)$ denote prior and posterior densities. Then we may write down the posterior of the number of clusters and latent cluster allocations from Bayes' theorem
\begin{eqnarray}
\pi(K,G,\bz,\bw,\boldomega,\boldrho,\Theta|Y)& \propto & p(\bz|K,\boldomega)p(\bw|G,\boldrho) p(Y|K,G,\bz,\bw,\Theta) \nonumber \\
& & \qquad \qquad \times \pi(\Theta|K,G)\pi(\boldomega|K) \pi(\boldrho|G) \pi(K) \pi(G).
\end{eqnarray}

Adopting a conjugate prior for $\boldomega$, $\boldrho$ and each $\theta_{kg}$ allows one to integrate these from the posterior analytically. We call this collapsing. Doing this allows us to obtain the marginal posterior $\pi(K,G,\bz,\bw|Y)$. Samples can be generated from this posterior using the MCMC sampler of Section~\ref{sec:mcmc}.  This is similar to the general approach of~\citeasnoun{Nobile07}. The idea of collapsing has been used by~\citeasnoun{Sheng03} in the analysis of a discretized microarray and by~\citeasnoun{Griffiths04} in latent Dirichlet analysis for document classification. It would be possible to estimate this model without integrating out parameters by using reversible jump MCMC (RJMCMC)~\cite{Green95}. We discuss this further in Section~\ref{sec:rjcomparison}.

We choose a standard conjugate prior for each of the parameters to be integrated out. For example, $\boldomega \sim \mbox{Dirichlet}(\alpha,\dots,\alpha)$ and $\boldrho\sim\mbox{Dirichlet}(\beta,\dots,\beta)$ {\it a priori}. For the examples considered in this paper, we take the non-informative values $\alpha=1,\beta=1$. The prior on $\theta_{kg}$ will depend on the distribution assumed for the data. For the most widely used models, a standard conjugate prior will be available. The $\theta_{kg}$ are assumed independent {\it a priori}. This leads to the posterior 
\begin{equation}
\pi(K,G,\bz,\bw|Y) \propto \pi(K) \pi(G) \frac{\Gamma\{\alpha K\}\prod_{k=1}^K \Gamma\{n_k+\alpha\}}{\Gamma\{\alpha\}^K \Gamma\{n+\alpha K\}}\frac{\Gamma\{\beta G\}\prod_{g=1}^G \Gamma\{m_g+\beta\}}{\Gamma\{\beta\}^G \Gamma\{m+\beta G\}}\prod_{k=1}^K \prod_{g=1}^G  M_{kg}. \label{eq:posterior_KG}
\end{equation}
where 
\[
M_{kg} = \int \pi(\theta_{kg}) \prod_{i:z_i=k} \prod_{j:w_j=g} p(y_{ij}|\theta_{kg}) \, \mbox{d} \theta_{kg}.
 \] 
where $n_k$ is the number of rows in cluster $k$ and $m_g$ is the number of columns in cluster $g$. An outline of the calculation of the posterior is given in Appendix A.

The priors for the number of clusters, $\pi(K)$ and $\pi(G)$ are taken to be truncated $\mbox{Poisson}(1)$ over the ranges $1,\dots,K_{\max}$ and $1,\dots,G_{\max}$. Examples of Poisson priors being adopted for the number of components include~\citeasnoun{Phillips96} and~\citeasnoun{Stephens00}. The use of a truncated $\mbox{Poisson}(1)$ prior has been justified in~\citeasnoun{Nobile05}. We did experiment with a uniform prior on the number of clusters. However, we found that this gave unnecessary empty clusters in some situations.

 We now give the $M_{kg}$ for two useful data models which we will use in examples later. 

\subsubsection{Bernoulli model for binary data} \label{Bernoulli}

Assume that $\Pr(y_{ij}=1|z_i=k,w_j=g)=\theta_{kg}$. We take a $\mbox{Beta}(\gamma,\delta)$ prior on $\theta_{kg}$. Then
\[
M_{kg} = \frac{\Gamma\{\gamma+\delta\}}{\Gamma\{\gamma\}\Gamma\{\delta\}} \frac{\Gamma\left\{ s_{kg}+\gamma\right\}\Gamma\left\{n_k m_g - s_{kg}+\delta\right\}}{\Gamma\{n_k m_g + \gamma+\delta\}}
\]
where the block sufficient statistic $s_{kg}=\sum\limits_{i:z_i=k}\, \sum\limits_{j:w_j=g} y_{ij}$. Further detail on the calculation is given in Appendix B.
\subsubsection{Gaussian model for continuous data} \label{Gaussian}

Assume $y_{ij}|z_i=k,w_j=g \sim \mbox{N}(\mu_{kg},\sigma_{kg}^2)$. Take the priors $\mu_{kg}  \sim  \mbox{N}(\xi,\tau^2\sigma_{kg}^2)$ and $\sigma_{kg}^2  \sim  \mbox{IG}(\delta/2,\gamma/2)$ where $\mbox{IG}(a,b)$ is the Inverse-Gamma distribution: $p(x) = \frac{b^a}{\Gamma\left(a\right)}x^{-(a+1)}\exp\{-b/x\}$. Then
\begin{eqnarray*}
M_{kg} = \frac{\gamma^{\delta/2}\,\Gamma\{(n_k m_g + \delta)/2\}}{\pi^{n_k m_g/2}\Gamma\{\delta/2\}\left(n_k m_g \tau^2 +1\right)^{1/2}}\left(ss_{kg} - \frac{\tau^2 \left(s_{kg} + \xi/\tau^2 \right)^2}{n_k m_g \tau^2 + 1}+ \frac{\xi^2}{\tau^2}+\gamma\right)^{-(n_k m_g + \delta)/2}
\end{eqnarray*}
where $s_{kg}=\sum\limits_{i:z_i = k} \, \sum\limits_{j:w_j=g} y_{ij}$ and $ss_{kg}=\sum\limits_{i:z_i = k} \, \sum\limits_{j:w_j=g} y_{ij}^2$. Further details on calculating $M_{kg}$ are given in Appendix B.

\section{MCMC sampling of clusterings} \label{sec:mcmc}

The sampler which we propose consists of four different moves. The first is just a standard Gibbs update for the row/column label. The second proposes to reallocate collections of rows and columns. The final two moves propose to add or remove clusters. We describe the moves for rows, but they apply to columns analogously. When running the algorithm, the moves are each applied to the rows and columns in a single sweep. Since the LBM will be invariant to cluster labellings, we will encounter the label switching problem. We outline how to deal with this as well as discussing how to summarize the output from the sampler.

\subsection{MCMC moves}

\subsubsection{Gibbs sampling to update the allocation of one row}

Suppose row $i$ is currently in cluster $k$. We then sample its new allocation, $\tilde{z}_i$ from the distribution  
\begin{equation}
p(\tilde{z}_i = k'|Y,K,G,\bz_{-i},\bw) \propto \frac{n_{k'}+\alpha}{n_k-1+\alpha} \prod_{g=1}^G \frac{ M_{k'g}^{(+i)} M_{kg}^{(-i)}}{M_{k'g}M_{kg}} \quad, k' \ne k\label{eq:gibbs}
\end{equation}
and $p(\tilde{z}_i=k|Y,K,G,\bz_{-i},\bw)\propto 1$ where  $M_{kg}^{(-i)}$ and $M_{k'g}^{(+i)}$ are obtained respectively by removing row $i$ from cluster $k$ and adding it to cluster $k'$ within column cluster $g$. The total computational effort required for the Gibbs sweep on rows and columns is $O((n+m)KG)$ which may be prohibitive for large $K$,$G$,$n$ or $m$. It is possible to move one row and column between clusters using a Metropolis-Hastings move. This could be alternated with a Gibbs update to reduce computational overhead or some mixture of the two moves could be used. 

\subsubsection{Move to update the allocation of more than one row} \label{move:move2}

This move is similar to move M3 in~\citeasnoun{Nobile07}. Its role is to move more than one row at a time. The way in which new row allocations are proposed should isolate clusters more quickly than just performing one row Gibbs updates. The procedure is as follows. Choose two row clusters $k$ and $k'$ at random. Let $S$ be the index set of rows currently belonging to clusters $k$ and $k'$. The members of $S$ are randomly reordered. Imagining clusters $k$ and $k'$ to be empty initially and $S$ to be full, we sequentially take each row from $S$ and allocate it to $k$ or $k'$. This allocation is done using the probability that the current clusters $k$ or $k'$ generated that row conditioning on rows that have already been reallocated to $k$ or $k'$. For row $i$ in $S$ these probabilities are denoted by $p_{k}^{(i)}$ and $p_{k'}^{(i)}$ with $p_{k}^{(i)}+ p_{k'}^{(i)}= 1$. To write down the proposal probability of this move we use $\bar{M}_{kg}, g=1,\dots,G$ to represent the integrated likelihood of the members placed in cluster $k$ before member $i$ has been processed. Similarly $\bar{n}_k$ represents the number of rows in cluster $k$ before $i$ has been processed. Then using similar notation to the Gibbs move it can be shown (see Appendix A.2,~\citeasnoun{Nobile07}) that
\[
\frac{p_{k'}^{(i)}}{p_{k}^{(i)}} = \frac{\bar{n}_{k'}+\alpha}{\bar{n}_k+\alpha} \prod_{g=1}^G \frac{\bar{M}_{k'g}^{(+i)}\bar{M}_{kg}}{\bar{M}_{k'g}\bar{M}_{kg}^{(+i)}}.
\] 
Using $p_{k}^{(i)}+ p_{k'}^{(i)}= 1$, the above can be solved for $p_{k'}^{(i)}$. The proposed allocation of row $i$, $\tilde{z}_i$ may then be sampled. Once the quantities $\bar{n}_{k'},\bar{n}_k,\bar{M}_{k'g}$ and $\bar{M}_{kg}$ have been updated based on $\tilde{z}_i$, the next row in $S$ can be dealt with.

When all members of $S$ have been processed the proposal probability of moving from $\bz$ to $\tilde{\bz}$ is
\[
\frac{1}{K(K-1)} \prod_{i \in S} p_{\tilde{z}_i}^{(i)}.
\]
For the reverse move the proposal probability is
\[
\frac{1}{K(K-1)} \prod_{i \in S} p_{z_i}^{(i)}. 
\]
The new allocation $\tilde{\bz}$ is then accepted with probability $\min(1,A)$ where
\[
A = \frac{\Gamma\{\tilde{n}_k + \alpha\}\Gamma\{\tilde{n}_{k'}+\alpha\}}{\Gamma\{n_k + \alpha\}\Gamma\{n_{k'}+\alpha\}}\prod_{g=1}^G  \frac{\tilde{M}_{k'g}\tilde{M}_{kg}}{M_{k'g}M_{kg}}\times \prod_{i\in S}  \frac{p_{z_i}^{(i)}}{p_{\tilde{z}_i}^{(i)}}.
\]
and $\tilde{n}_k,\tilde{n}_{k'},\tilde{M}_{k'g},\tilde{M}_{kg}$ are the proposed cluster sizes and integrated block likelihoods when all the members of $S$ have been processed.

\subsubsection{Moves to split or combine clusters} \label{move:move3}

To add a cluster we first randomly propose a cluster, $k$, to ``split''. The new cluster will be labelled $K+1$ if the current number of clusters is $K$. In the same way as the move to reallocate more than one row (Section~\ref{move:move2}), the probability of a row proposed as being in cluster $k$ or $K+1$ is given by the conditional probability it was generated by that cluster, the rows being processed sequentially. Clearly the order in which rows are processed is important. Thus for the split and combine moves we place an ordering on the members of cluster $k$, that is, the order in which the members are arranged in cluster $k$ is important. As well as taking members out from cluster $k$ and placing them in cluster $K+1$, this is important when we place all members back into cluster $k$ in the combine move. It is possible to propose a label swap of $K+1$ with any other label selected at random (itself included), say $k'$. This then would split cluster $k$ into clusters $k$ and some $\{1,\dots,K+1\}\backslash k$. 

Let $S$ denote the index set of rows currently belonging to cluster $k$. We choose a split move with probability $p_{\mbox{s}}^K$. For the split move, the denominator in the proposal ratio will be
\[
p(\bz \rightarrow \tilde{\bz}) = p_{\mbox{s}}^K \frac{1}{K(K+1)}\frac{1}{n_k!}\prod_{i\in S} p_{\tilde{z}_i}^{(i)}.
\]
where the second term accounts for selecting the cluster to split, and then the cluster to swap labels with, the third term accounts for the number of ways in which members may be arranged (processed), and the fourth term is the product of conditional probabilities (see Section~\ref{move:move2}). 

For the combine move, two clusters are selected at random, say $k$ and $k'$ from the $K+1$ available. Then all members of cluster $k'$ are proposed to be placed back in $k$. Thus the numerator in the proposal probability for the split move is
\[
p(\tilde{\bz} \rightarrow \bz) = (1-p_{\mbox{s}}^{K+1})\frac{1}{K(K+1)} \frac{1}{n_k!}
\]
where the first term is the probability of proposing a combine move, the second accounts for the clusters selected, and the third accounts for the number of ways in which the members of cluster $k$ may be arranged.

The acceptance probability for the split move is then $\min(1,A)$ where
\begin{eqnarray*}
A &=& \frac{\pi(K+1)}{\pi(K)} \frac{\Gamma\{n + \alpha K\}}{\Gamma\{n+\alpha(K+1)\}} \frac{\Gamma\{\alpha(K+1)\}}{\Gamma\{\alpha\}\Gamma\{\alpha K\}} \frac{\Gamma\{\tilde{n}_k + \alpha\}\Gamma\{\tilde{n}_{k'}+\alpha\}}{\Gamma\{n_k+\alpha\}}\\
& &  \qquad \qquad \qquad \times \prod_{g=1}^G \frac{\tilde{M}_{kg} \tilde{M}_{k'g}}{M_{kg}} \times \frac{1-p_{\mbox{s}}^{K+1}}{p_{\mbox{s}}^K} \left(\prod_{i\in S} p_{\tilde{z}_i}^{(i)}\right)^{-1}
\end{eqnarray*}
and $\tilde{n}_k ,\tilde{n}_{k'},\tilde{M}_{kg},\tilde{M}_{k'g}$ give the proposed sizes and integrated block likelihoods of the proposed clusters. 

The acceptance probability for the combine move is $\min(1,A^{-1})$. These moves are similar to the ``split and combine'' moves discussed by~\citeasnoun{Richardson97}. Our experiments showed that they gave satisfactory mixing and higher acceptance rates than proposing empty clusters.

\subsection{Form of reversible jump sampler}\label{sec:rjcomparison}

As noted, the moves discussed in the previous section resemble moves used in the RJMCMC sampler of~\citeasnoun{Richardson97} and other RJ samplers for related classification problems (for example~\citeasnoun{Robert00}). The difference with our sampler is that the space we sample from is of fixed dimension. This is due to collapsing. Performing an equivalent RJ analysis to that presented here would be challenging for LBMs. This would mean extending the Gibbs sampler of~\citeasnoun{vanDijk09} to include variable dimension moves for splitting or combining clusters. The construction of proposal densities for variable dimensional moves in RJ samplers can be crucial to their performance. Work has been done in this area~\cite{Brooks03,Green03}, but for many applications construction of proposals is case specific. The reason for entertaining a RJ analysis here is that we are not only concerned with finding a cluster allocation for a specified LBM. We are also interested in exploring different cluster models, and so the task also becomes one of model determination as discussed in Section~\ref{sec:determine_k_and_g}.

Consider splitting a row cluster $k$ into $k$ and $k'$ in a typical RJ approach. This is more difficult than the component splitting case in~\citeasnoun{Richardson97}, since splitting each row cluster gives rise to $d(G+1)$ new parameters where $d$ is the dimension of any $\theta_{kg}$. Finding a proposal that will mix well may require lots of trial and error, especially if $d>1$ or $G$ is even moderately large. Moreover, computational time would increase dramatically with respect to the collapsed LBM in these situations. 

Using a collapsed model, is, in a sense, a form of variance reduction for this model. We reduce variability in sampling of allocations, by integrating out $\boldomega,\boldrho$ and $\Theta$. This should give better sampling of the high probability clusterings of the data, since uncertainty due to parameter values has vanished.

\subsection{Label switching}

The joint posterior of cluster models and allocations or labels (\ref{eq:posterior_KG}) is invariant to label switching, that is, the labels are not identifiable. If there is one labelling $1,\dots,K$ of the rows, then any permutation of this, say, $\sigma(1),\dots,\sigma(K)$, gives exactly the same information about clustering relationships. The posterior on row labels has $K!$ indistinguishable modes. Generally as the Markov chain progresses, we will observe switches between these equivalent modes; the well known label switching phenomenon. In our case label switching can occur for row and column labels independently. There are many approaches for dealing with the label switching problem~\cite{Stephens00,Celeux00}. The approach we adopt here is due to~\citeasnoun{Nobile07}. It is ideal for our purposes since it does not involve loss functions based on sampled model parameters (which are no longer in our model). It just requires the samples of $\bz$ and $\bw$. 

We now outline the procedure we use to deal with label switching. Some more details are given in Appendix C. We post-process row and column allocation vectors separately. The re-labelled data can then be used to compute posterior probability of cluster memberships and other quantities of interest. To post process the label vectors $\bz_1,\bz_2,\dots$ output from MCMC we begin by arranging these in order of increasing number of non-empty components. This gives the ordering $\bz^{(1)},\bz^{(2)},\dots$, where for $s<t$, $\bz^{(s)}$ uses either the same number of components as $\bz^{(t)}$ in total, or less. For example, with $K=4$, $\bz^{(s)}=(3,3,2,2,2,1)$ would come before $\bz^{(t)} = (4,4,3,3,1,2)$. Suppose we have processed and re-labelled the vectors $\bz^{(t)}$ up to time $T-1$, and there are $K_{T-1}$ non-empty components in $\bz^{(T-1)}$. Compute a cost matrix with general element
\[
C(k_1,k_2) = \sum_{t=1}^{T-1} \sum_{i=1}^n \mbox{I}\left\{z_i^{(t)} \ne k_1, z_i^{(T)} = k_2 \right\}.
\]
Then the more $\bz^{(T)}$ disagrees with the vectors already processed, the higher this cost will be (see Appendix C). The square assignment algorithm of~\citeasnoun{Carpaneto80} returns the permutation $\sigma(\cdot)$ of the labels in $\bz^{(T)}$ which minimizes the total cost $\sum_{k=1}^{K_{T-1}} C(k,\sigma(k))$. We then relabel $\bz^{(T)}$ by permuting the labels according to $\sigma(\cdot)$.

\subsection{Summarizing MCMC output} \label{sec:estimated}

Having sampled both the number of clusters and cluster memberships, it will be of interest to give a summary of the sampling. As different $(K,G)$ cluster models are structurally different, it is not possible to give an ``average'' of cluster membership. We suggest two summaries. 

\subsubsection{Using the modal cluster model} \label{sec:modal_model}
The first summary focuses on using the modal, or most visited model from the MCMC output. It takes the series of $(K,G)$ visited models and chooses the pair which appear most often. Call this pair $(\hat{K},\hat{G})$. Suppose this pair has occured $N$ times in the post burn-in sample. We extract the $N$ pairs of label vectors $\bz$ and $\bw$ corresponding to these occurrences. We then post process these label vectors using the procedure to undo label switching outlined in the previous section. This will be necessary to compute posterior distributions of row and column cluster membership. After computing the posterior distributions of row membership, row $i$ has distribution $(q_{i1},\dots,q_{i\hat{K}})$ where $q_{ik}$ is the estimated posterior probability row $i$ belongs to cluster $k$ in the $(\hat{K},\hat{G})$ cluster model. For the summary we assign $i$ to cluster $\arg \max_k q_{ik}$. The columns are given the same treatment.

\subsubsection{Using the MAP}
Since we are sampling from the fixed dimensional posterior $\pi(K,G,\bz,\bw|Y)$, the maximum {\it a posteriori} (MAP) cluster model and cluster membership $(K,G,\bz,\bw)_{\mbox{\tiny MAP}}$ is also a useful summary of the MCMC output. The MAP gives the visited $(K,G,\bz,\bw)$ having highest probability {\it a posteriori} from the samples obtained.

\section{Simulation experiment}

\begin{figure*}
\begin{center}
$
\begin{array}{c}
\includegraphics[width=130mm]{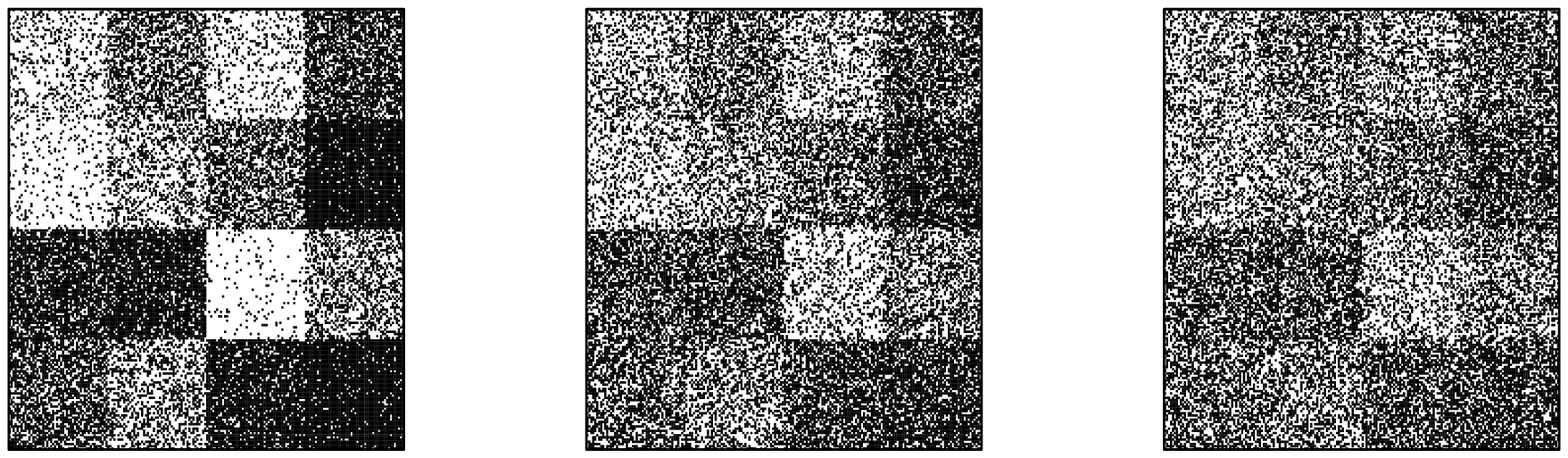}\\
\includegraphics[width=130mm]{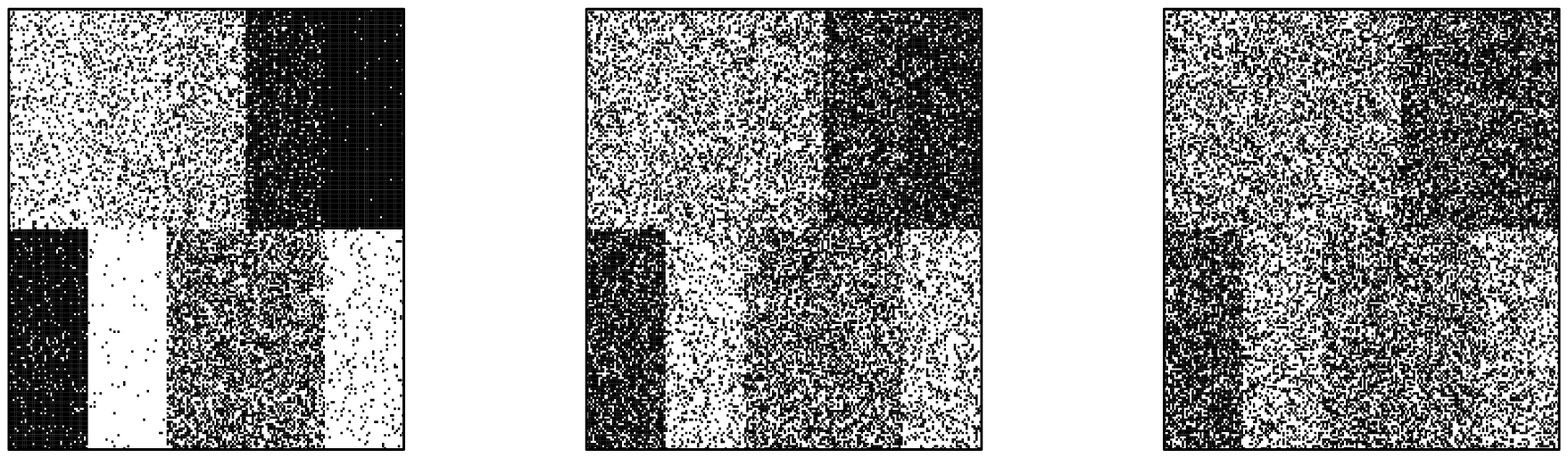}\\
\includegraphics[width=130mm]{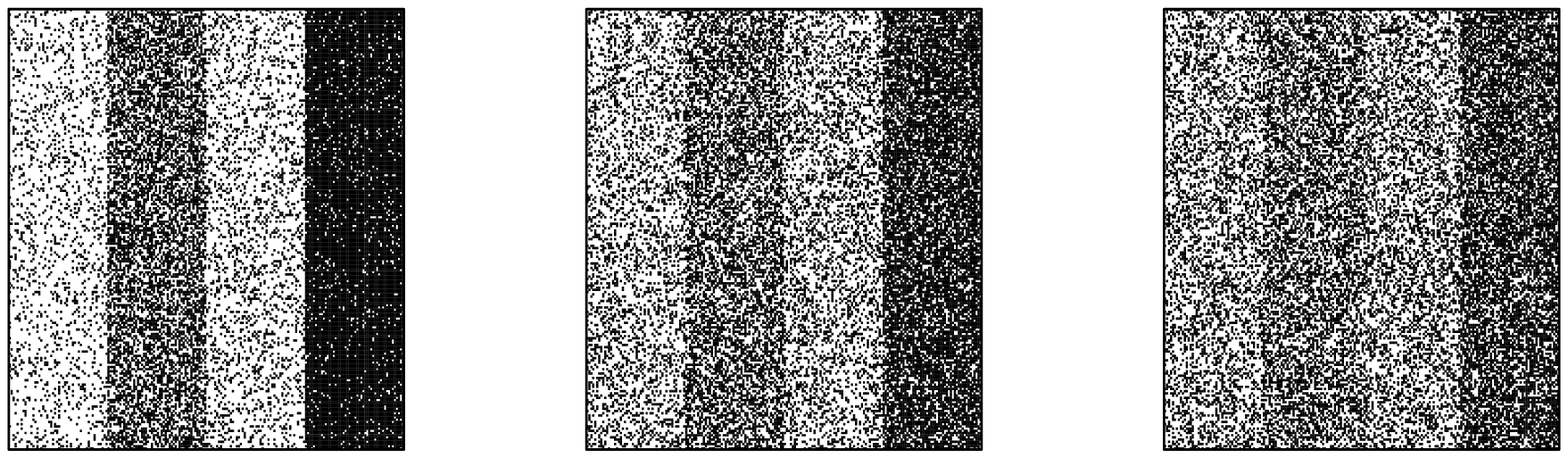}
\end{array}
$
\caption{Simulated data with decreasing distinguishability} \label{fig:sim_2}
\end{center}
\end{figure*}

To see how the sampler discriminates between different cluster models, it was run on some simulated data. We generated three $200 \times 200$ binary matrices with 4 and 4, 2 and 5 and 1 and 4 row and column clusters respectively. In each case, the block parameter $\theta_{kg}$ was drawn uniformly from $[0,1]$. The blocks were then generated using $\mbox{Bernoulli}(\theta_{kg})$ random variates. This is shown in the left of Figure~\ref{fig:sim_2}. Clusters were made less distinguishable by transforming the generated $\theta_{kg}$ to the intervals $[0.2,0.8]$ and $[0.3,0.7]$ using $\theta_{kg}^{[a,b]} = a + \theta_{kg}(b-a)$ and generating two further matrices. The rows and columns of the resulting matrices were then randomly reordered, disguising the data structure. The chain was run for 1000 burn-in iterations and a further 16,000 iterations on each data set. We assumed a $\mbox{Beta}(1,1)$ prior for $\theta_{kg}$ in all cases. The priors for $\boldomega$ and $\boldrho$ are as in Section~\ref{sec:BayesLBM}.

We looked at two performance diagnostics of the sampler. The first was the posterior model probability (PMP) of the model used to generate the data and the second was the integrated autocorrelation time (IAT) of sampled cluster models. Computing the PMP just amounts to counting the number of times the model in question was visited and dividing by the total number of samples. For the IAT we identify cluster models $(K,G)$ by a model index $R = 1,\dots,K_{\max}G_{\max}$. Then we estimate the quantity $\tau = 1+ 2\sum_{t=1}^{\infty} \rho_R(t)$, where $\rho_R(t)$ is the autocorrelation of the series of post burn-in samples $R_1,R_2,\dots$ at lag $t$. The series here refers to the cluster models sampled from the posterior (\ref{eq:posterior_KG}). Lower values of the IAT indicate better mixing and better performance of the MCMC sampler. The IAT can thus be used as a measure of efficiency for MCMC algorithms. See for example~\citeasnoun{Liu04}, Chapter 5 and~\citeasnoun{Roberts96}.

\begin{table}
\begin{center}
\begin{tabular}{l|cccc|}
\hline
$(K,G)$ & $\theta_{kg}$ & PMP & $\hat{\tau}$\\
\hline
(4,4) & A & 0.9550 & 8.79 \\ 
 			& B & 0.9463 & 10.57 \\
 			& C & 0.9014 & 17.43 \\
\hline
(2,5) & A & 0.9343 & 4.55 \\
 & B & 0.8886 & 9.79 \\
  & C & 0.8369 & 13.66 \\
\hline
 (1,4) & A & 0.8035 & 7.86 \\
  & B & 0.3000 & 8.97 \\
  & C & 0.1494 & 4.61 \\
\hline
\end{tabular}
\end{center}
\caption{Results of simulation experiment. The PMP gives the posterior model probability of the generating model. $\hat{\tau}$ is the estimated IAT.}\label{Tsimulated_ex} 
\end{table}

The results are shown in Table~\ref{Tsimulated_ex}. The $\theta_{kg}$ column is coded A for $\theta_{kg}\sim \mbox{Uniform}[0,1]$, B for transformation to $[0.2,0.8]$ and C for $[0.3,0.7]$. As the noise in the data increases, the ability to identify the model which generated the data decreases. This is to be expected. The estimated IAT indicates that we get less efficient sampling as the noise increases, with the exception of the $(1,4)$ cluster model. This is a particularly challenging situation, since two of the clusters are very similar (see Figure~\ref{fig:sim_2}). The transformed $\theta_{11},\theta_{13}$ were for $[0.2,0.8]:0.288,0.320$ and for $[0.3,0.7]:0.36,0.38$. The fact that these two clusters are practically indistinguishable would make the sampler choose a $(1,3)$ model as the best model after scrambling of the data. In fact the most visited model in both these cases had 1 row cluster and 3 column clusters (62.37\% and 82.66\% of the posterior probability). In this situation, the best cluster model was not the same as the generating model. This is an artifact of the simulation process but shows that a sensible clustering can be achieved.

\section{Congressional voting in US senate} \label{sec:congressional_voting}

We apply the sampler to the UCI Congressional Voting data assuming the Bernoulli model of Section~\ref{Bernoulli}. The data records whether 435 members of the $98^{\mbox{\tiny{th}}}$ congress (267 democrats, 168 republicans) voted ``yay'', ``nay'', abstained or were absent in votes on 16 different key issues. Here the members of congress are represented by rows, and the issues are represented by columns. The data is available from

\begin{figure}
\begin{center}
\includegraphics[width=150mm]{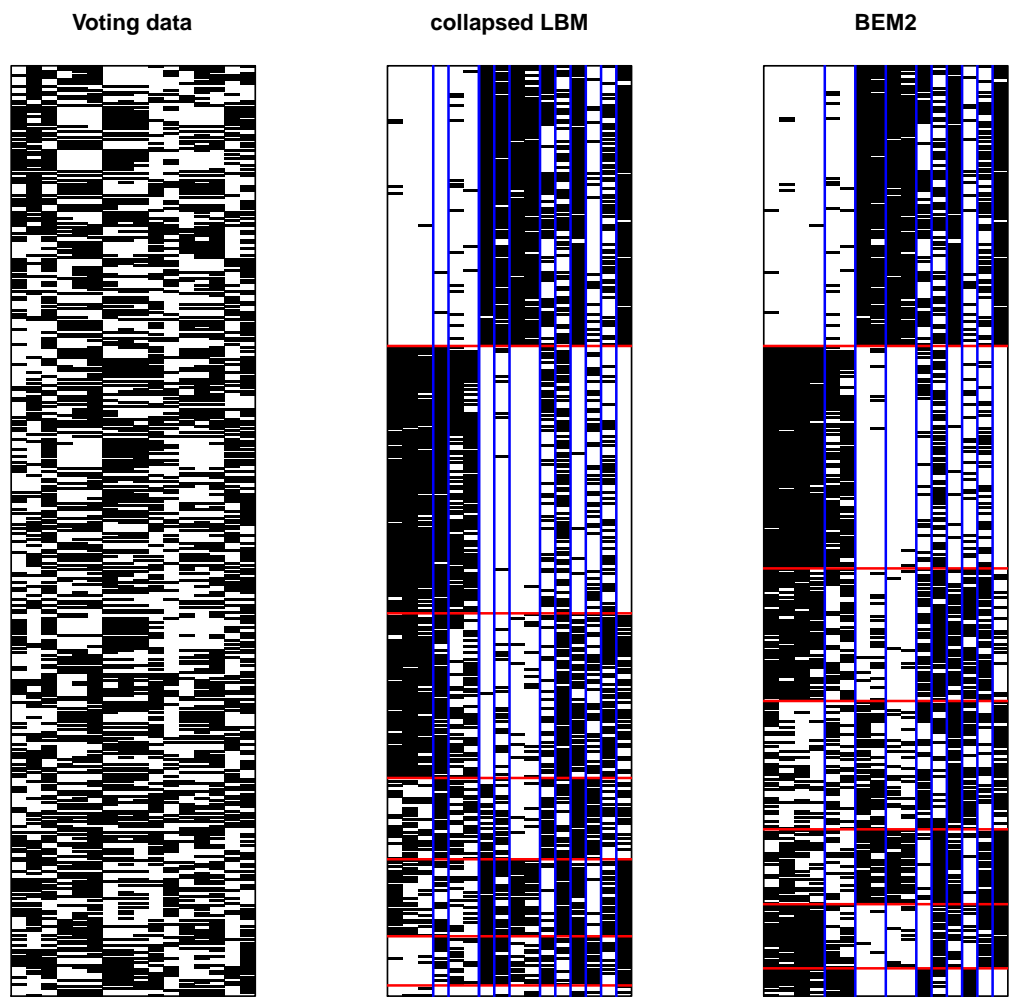}
\end{center}
\caption{Voting data. Colour key: white = ``nay'',black = ``yay''. Left panel: Raw data. Right panels: summary cluster membership from the modal 7 row and 12 column cluster model and the cluster membership obtained from BEM2. Row clusters are numbered 1-7, top-bottom. Column clusters are numbered left-right. The red lines divide clusters of congressmen, and the blue lines divide the issue clusters.}\label{fig:cong_clust}
\end{figure}

 \texttt{http://archive.ics.uci.edu/ml/datasets/Congressional+Voting+Records}
 
\noindent and is shown in the left panel of Figure~\ref{fig:cong_clust}. The aim was to see whether the sampler could discover any clustering by party and issue. For example, one may expect that democrats voted differently to republicans on certain issues. It was thought best to ignore absent and abstain votes. Here, this is equivalent to treating these votes as a ``nay'', since our focus is on clustering rows and columns. The only sample sizes entering into our calculations are the number of rows and columns in each cluster. For the Bernoulli model, the block sufficient statistic is the sum of the data. This is not affected by a missing data point.
We do point out however that missing data could be easily imputed by inclusion of a Gibbs step to sample from the full conditional of any missing point.

The sampler was run for 110,000 iterations with 10,000 as a burn-in initialized at the no cluster model. To reduce correlation in samples, we took every $10^{\mbox{th}}$ sample after burn-in. The move of Section~\ref{move:move2} had a 16\% acceptance rate for rows and 51\% for columns. The cluster split and combine moves had about a 1.5\% acceptance rate for rows and 8\% for columns. The run took just over an hour on a 2.5GHz processor.

Table~\ref{tab:dist_cong} shows the distribution of the number of row and column clusters. It can be seen that about 60\% of the posterior probability is placed on 6/7 row clusters and 12/13 column clusters. We extracted the samples with 7 row clusters and 12 column clusters to construct an estimated clustering following Section~\ref{sec:modal_model}. The estimated clustering is shown in the middle panel of Figure~\ref{fig:cong_clust}. The red horizontal lines here divide the clusters of congressmen, and the blue vertical lines divide the issue clusters. When referencing Figure~\ref{fig:cong_clust} we say that the congressman clusters (rows) are numbered 1 to 7 top-bottom, and the issue clusters (columns) are numbered 1 to 12 left-right.

Issues have only three non-singleton clusters. The first contains ``anti-satellite-test-ban'', ``aid-to-nicaraguan-contras'' and ``mx-missile'' (column cluster 1). The second contains ``physician-fee-freeze'' and ``education-spending'' (column cluster 3) and the third has ``handicapped-infants'' and ``duty-free-exports'' (column cluster 6). Row cluster composition by party is shown in Table~\ref{tab:split}. The majority democrat party roughly splits into four clusters, while the republican party splits into two. The main discrepancy between the two large democrat clusters, 2 and 3, appear to be the issues ``religious-groups-in-schools'' and ``crime'' in issue clusters 9 and 12. Row cluster 6 which is also mainly democrat appears to vote similarly to the republican cluster 1. Row clusters 4 (democrat) and 5 (republican) appear to deviate from their core party vote.

\begin{table}
\begin{centering}
\begin{tabular}{ccccccccc}
  Columns &   9   &  10  &   11  &   12 &    13  &   14  &   15  &   16 \\
  Rows & & & & & & & & \\
   5 & 0.0000 & 0.0002 & 0.0013 & 0.0008 & 0.0005 & 0.0000 & 0.0002 & 0.0000\\
   6 & 0.0001 & 0.0021 & 0.0401 & 0.1238 & 0.1491 & 0.0946 & 0.0348 & 0.0029\\
   7 & 0.0001 & 0.0041 & 0.0543 & 0.1675 & 0.1614 & 0.0697 & 0.0188 & 0.0012\\
   8 & 0.0000 & 0.0009 & 0.0101 & 0.0270 & 0.0191 & 0.0075 & 0.0029 & 0.0005\\
   9 & 0.0000 & 0.0000 & 0.0010 & 0.0013 & 0.0013 & 0.0007 & 0.0001 & 0.0000\\
\end{tabular}
\end{centering} \caption{Distribution of cluster models for voting data} \label{tab:dist_cong}
\end{table}

\begin{table}[h]
\begin{center}
\begin{tabular}{lcc}
\multicolumn{1}{c}{\bf Cluster}  & Democrat  &Republican\\
\hline \\
{\bf 1} (131) & 8 & 123  \\
 {\bf 2} (125)& 125 & 0 \\
 {\bf 3} (77) & 71 & 6\\
{\bf 4} (38) & 37 & 1  \\
{\bf 5} (36) & 3 & 33  \\
{\bf 6} (23) & 21 & 2  \\
{\bf 7} (5) & 2 & 3  \\
\end{tabular}
\end{center} \caption{Party distribution over row clusters from collapsed sampling.} \label{tab:split}
\end{table}

We compared the results obtained from our algorithm with those obtained from the BEM2 algorithm of~\citeasnoun{Govaert08}, reviewed in Section~\ref{sec:EM_explanation}. The algorithm was run using 7 row clusters and 12 column clusters. It should be noted that BEM2 requires the number of row and column clusters to be assumed known in advance. To obtain an estimated clustering, we took the cluster with the maximum probability of membership. The composition by party of the row clusters from BEM2 is shown in Table~\ref{tab:split2}. Row cluster 1 is similar in both, but there are some differences in the other clusters. The BEM2 clustering only used 10 column clusters of the 12 available. The right panel of Figure~\ref{fig:cong_clust} shows the clustering from BEM2. For comparison purposes with the collapsed LBM clustering, the columns have been arranged in the same order. The collapsed LBM appears to identify more small clusters, leading to a marginally more homogeneous blocking of the data. 

\begin{table}[h]
\begin{center}
\begin{tabular}{lcc}
\multicolumn{1}{c}{\bf Cluster}  & Democrat  &Republican\\
\hline \\
{\bf 1} (131) & 8 & 123  \\
 {\bf 2} (104)& 104 & 0 \\
 {\bf 3} (62) & 61 & 1\\
{\bf 4} (60) & 50 & 10  \\
{\bf 5} (35) & 5 & 30  \\
{\bf 6} (30) & 26 & 4  \\
{\bf 7} (13) & 13 & 0  \\
\end{tabular}
\end{center} \caption{Party distribution over row clusters from BEM2 algorithm.} \label{tab:split2}
\end{table}

There is an advantage here over an EM approach to fitting a cluster model in that the number of clusters need not be assumed known in advance. Here, model uncertainty is naturally inbuilt into the approach and it is dealt with automatically by the sampler. There is no user intervention to choose the cluster model. User intervention is only in the choice of prior hyperparameters and priors on the number of clusters. In our experience, standard non informative priors for the data model parameters, and the truncated Poisson(1) prior on the number of clusters (as argued in~\citeasnoun{Nobile05}), seem to perform well. The computations for the collapsed LBM are also numerically stable if clusters empty out. In our experimentation with EM algorithms this caused instability. Empty clusters could easily occur, say, if the chosen cluster model is not well supported by the data.

\section{Microarray experiments} \label{sec:microarray_experiments}

A DNA microarray experiment records expression levels of a large number of genes over a number of conditions or samples. The number of conditions or samples is usually less than 100, while the number of genes could be in the thousands. Discovering which genes behave similarly and under which subgroups of conditions is the aim of analysis. One way to do this is to group together genes with similar expression levels. Methods differ in whether they allow clusters to overlap or not. Here we will not allow clusters to overlap due to the form of the LBM. 

Analyzing DNA experiments can be challenging, due to the large row dimension and the general uncertainty in how many clusters may be present in the data. We apply our sampler to data from DNA experiments on the budding yeast {\it Saccharomyces Cerevisiae}. The microarray contains 419 genes and records the expression level of these under 70 conditions. It was obtained from the R package \texttt{biclust}~\cite{biclust09}. Expression levels lay between $-6$ and $+7$. The aim is to see how much structure the sampler can uncover, so the rows and columns of the microarray were randomly reordered (Figure~\ref{fig:bicat_clust} (a)). In our application the rows represent the genes and the columns represent the conditions.

We use the Gaussian model of Section~\ref{Gaussian} for expression level. This model requires specification of four hyperparameters. Two of these ($\gamma$ and $\delta$) are for the prior on the block error variances and two are for the prior on the block means ($\xi$ and $\tau^2$). We choose $\gamma = \delta = 0.02$ and $\xi = 0, \tau^2 = 100$. This choice of $\gamma$ and  $\delta$ gives a proper density on the error variance which is non-informative (see for example~\citeasnoun{Spiegelhalter96}). Similarly, choosing $\xi=0$ is a reasonable non-informative choice given the range of the data. Setting $\tau^2=100$ says that the prior information on a block mean is equal to 1\% of the information in the observed expression level of one gene under one condition within that block. This is also non-informative.

The sampler was run for 220,000 iterations with 20,000 taken as burn-in. We stored every $20^{\mbox{th}}$ iteration thereafter. The run was time consuming, taking approximately 3 hours. This said, the large gene dimension of such an array does pose a challenge when searching for two way clusters. The initial cluster model assumed had 1 row and column cluster $\mbox{i.e.}$ no cluster structure. Acceptance rates for the move of Section~\ref{move:move2} were 25\% for rows and 18\% for columns. Split and combine acceptances were about 0.5\% for rows and about 25\% each for columns. The low acceptance rates of split and combine moves for rows would be expected since finding clusters will be more difficult in a larger dimension.

Table~\ref{tab:microarray_dist} gives the PMP of the visited models from the MCMC output. The model space visited by the sampler is large. The modal model (25 row clusters and 4 column clusters) gave a posterior probability of 11.98\%. There is posterior support for anything from 3 to 5 column clusters, for 23 to 26 row clusters ($\mbox{PMP}>0.02$ in all cases). This is an example with considerable model uncertainty and it may be difficult to know the models to include using an information criterion over a grid of possible models as discussed in Section~\ref{sec:determine_k_and_g} and adopted by~\citeasnoun{vanDijk09}. Our approach has the obvious advantage of exploring the uncertainty in the posterior model space and attaching a probability to each model. 

Instead of constructing a summary based on the modal model, we took the MAP clustering here (shown in Figure~\ref{fig:bicat_clust} (b)). The MAP had 26 row clusters and 4 column clusters. 
To have a closer look at the row clusters, we plot a selection of these in Figure~\ref{fig:bicat_clust}. The plots show the gene expression profiles for genes in the same row cluster over conditions arranged by condition cluster. It can be seen that in certain cases, there is a clear clustering of genes with similar profiles. The gene clusters shown are arranged by size (left-right, top-bottom). Some of the larger clusters appear quite noisy, while some follow a common trend closely. 
Auxiliary runs of the sampler on the subsets of row clusters could be performed to try and isolate further cluster structures.

\begin{figure}
\begin{center}
$
\begin{array}{cc}
\includegraphics[width=30mm,height=90mm]{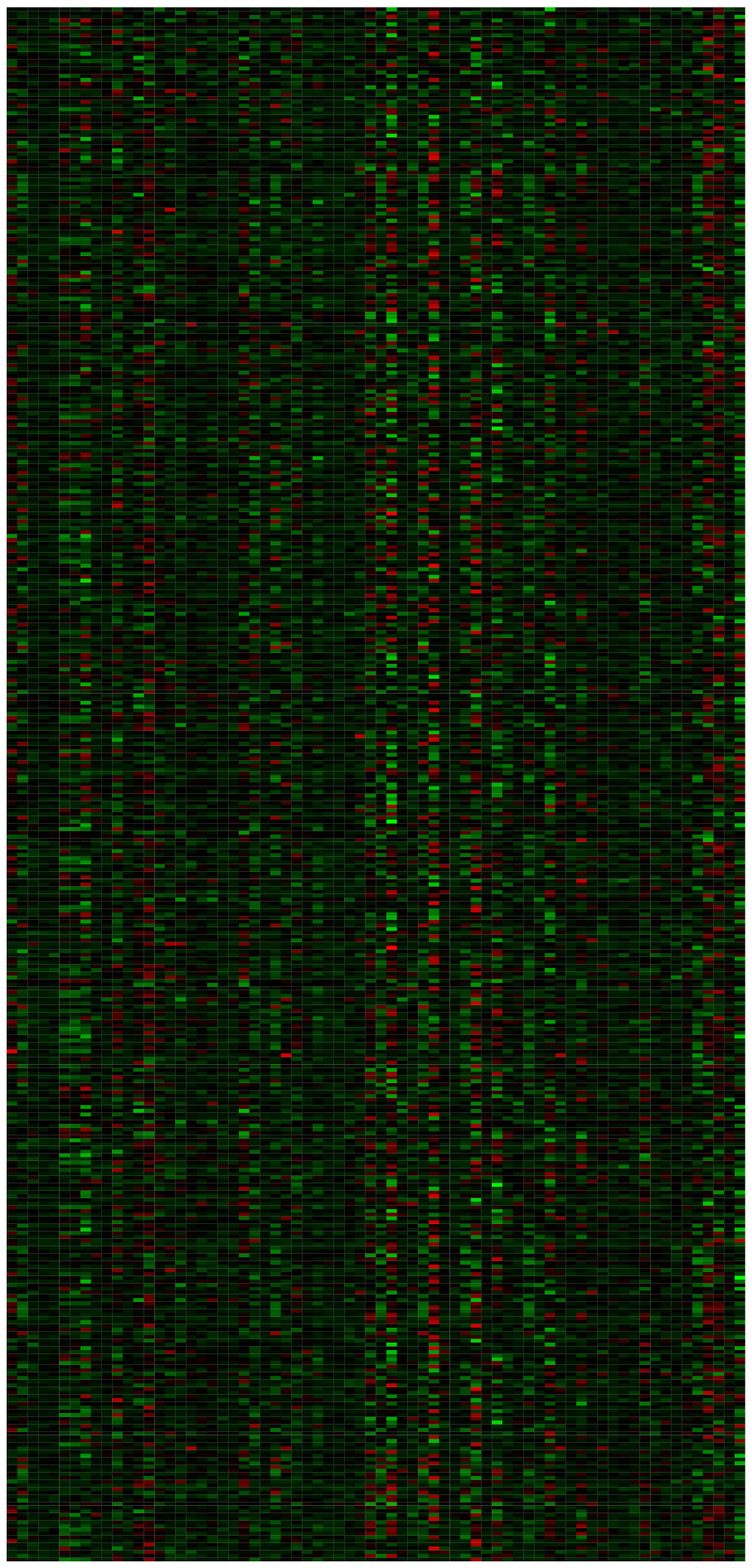} & \includegraphics[width=30mm,height=90mm]{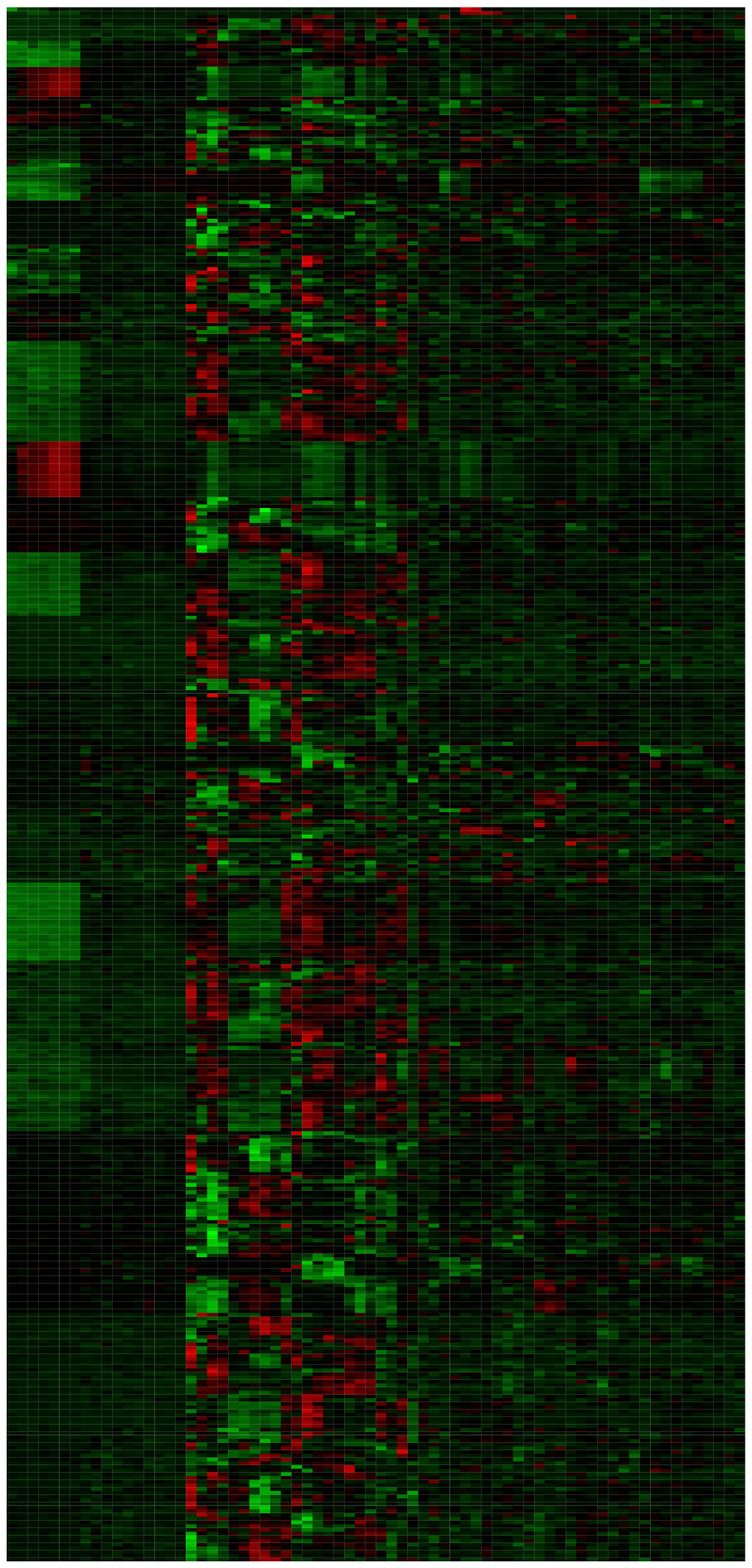} \\
\mbox{(a)} & \mbox{(b)}
\end{array}
$
\end{center}
\caption{Yeast data. (a) Original microarray (b) MAP clustering from sampler}\label{fig:bicat_clust}
\end{figure}

\begin{table}
\begin{center}
\small
\begin{tabular}{cccccccccccc}
    Columns  & 3 &      4   &   5  &    6 &      7    &  8    &  9   &  10   &  11 &   12  &  13\\
Rows  & & & & & & & & & & &   \\
  18 & 0.0000 & 0.0001 & 0.0000 & 0.0001 & 0.0000 & 0.0000 & 0.0000 & 0.0000 & 0.0000 & 0.0000 & 0.0000 \\
  19 & 0.0005 & 0.0002 & 0.0005 & 0.0001 & 0.0019 & 0.0044 & 0.0013 & 0.0011 & 0.0006 & 0.0000 & 0.0000 \\
  20 & 0.0005 & 0.0000 & 0.0002 & 0.0032 & 0.0110 & 0.0192 & 0.0121 & 0.0057 & 0.0012 & 0.0004 & 0.0000 \\
  21 & 0.0029 & 0.0027 & 0.0019 & 0.0023 & 0.0128 & 0.0147 & 0.0090 & 0.0028 & 0.0008 & 0.0003 & 0.0001 \\
  22 & 0.0076 & 0.0119 & 0.0074 & 0.0053 & 0.0109 & 0.0129 & 0.0085 & 0.0033 & 0.0007 & 0.0000 & 0.0001\\
  23 & 0.0368 & 0.0409 & 0.0242 & 0.0094 & 0.0089 & 0.0041 & 0.0030 & 0.0015 & 0.0002 & 0.0000 & 0.0000\\
  24 & 0.0643 & 0.0706 & 0.0420 & 0.0144 & 0.0073 & 0.0049 & 0.0023 & 0.0005 & 0.0002 & 0.0000 & 0.0000\\
  25 & 0.1016 & 0.1198 & 0.0702 & 0.0272 & 0.0082 & 0.0018 & 0.0011 & 0.0003 & 0.0002 & 0.0000 & 0.0000 \\
  26 & 0.0373 & 0.0457 & 0.0228 & 0.0095 & 0.0020 & 0.0008 & 0.0003 & 0.0000 & 0.0000 & 0.0000 & 0.0000\\
  27 & 0.0090 & 0.0109 & 0.0058 & 0.0021 & 0.0008 & 0.0002 & 0.0000  &0.0000 & 0.0000 & 0.0000 & 0.0000\\
  28 & 0.0008 & 0.0015 & 0.0010 &  0.0002 & 0.0002 & 0.0000 & 0.0000  & 0.0000 & 0.0000 & 0.0000&  0.0000
\end{tabular}
\normalsize
\end{center}
\caption{Posterior distribution of cluster models for the microarray data.} \label{tab:microarray_dist}
\end{table}

\begin{figure}
\begin{center}
$
\begin{array}{ccc}
\includegraphics[width=30mm,height=20mm]{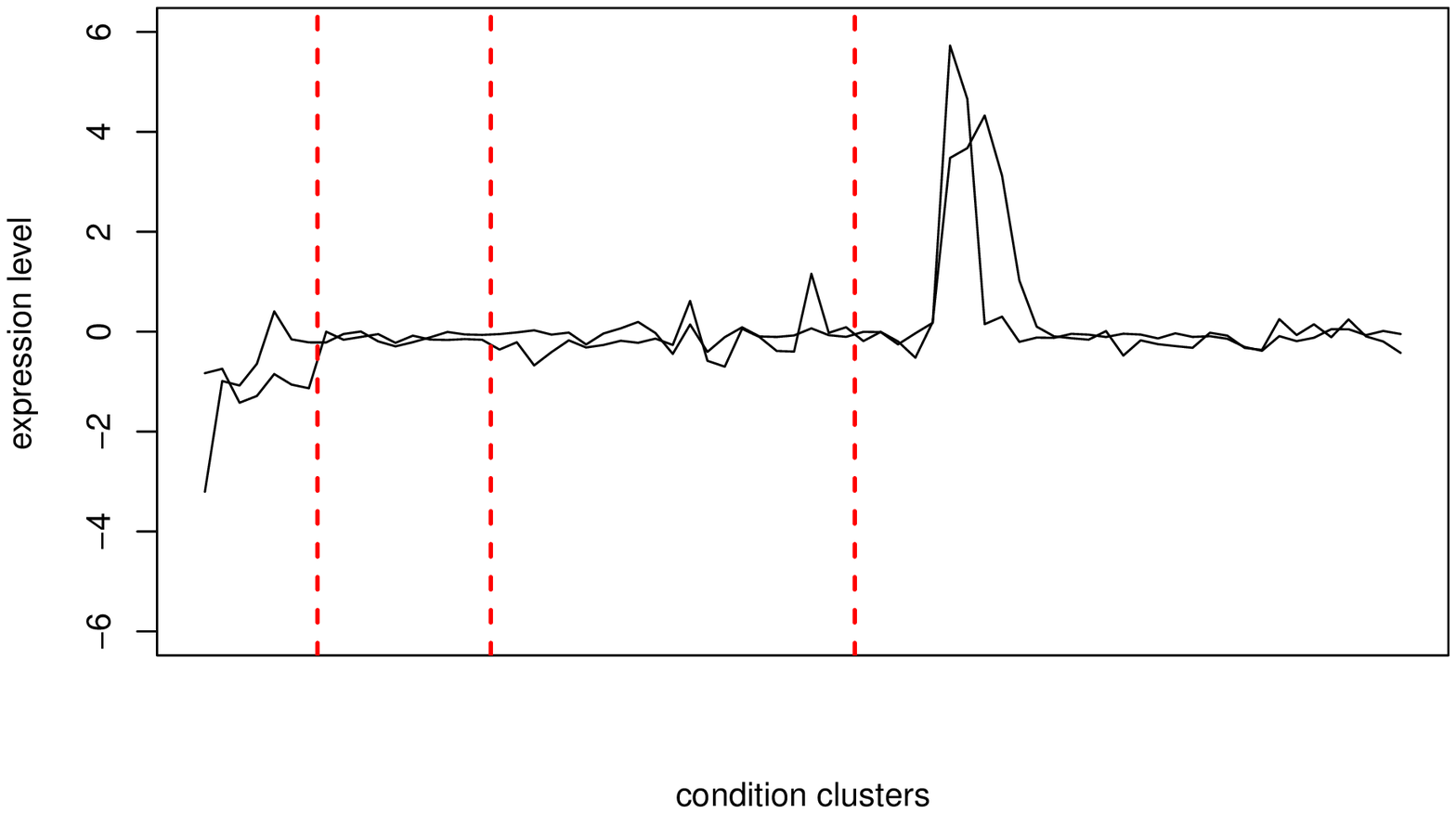} & \includegraphics[width=30mm,height=20mm]{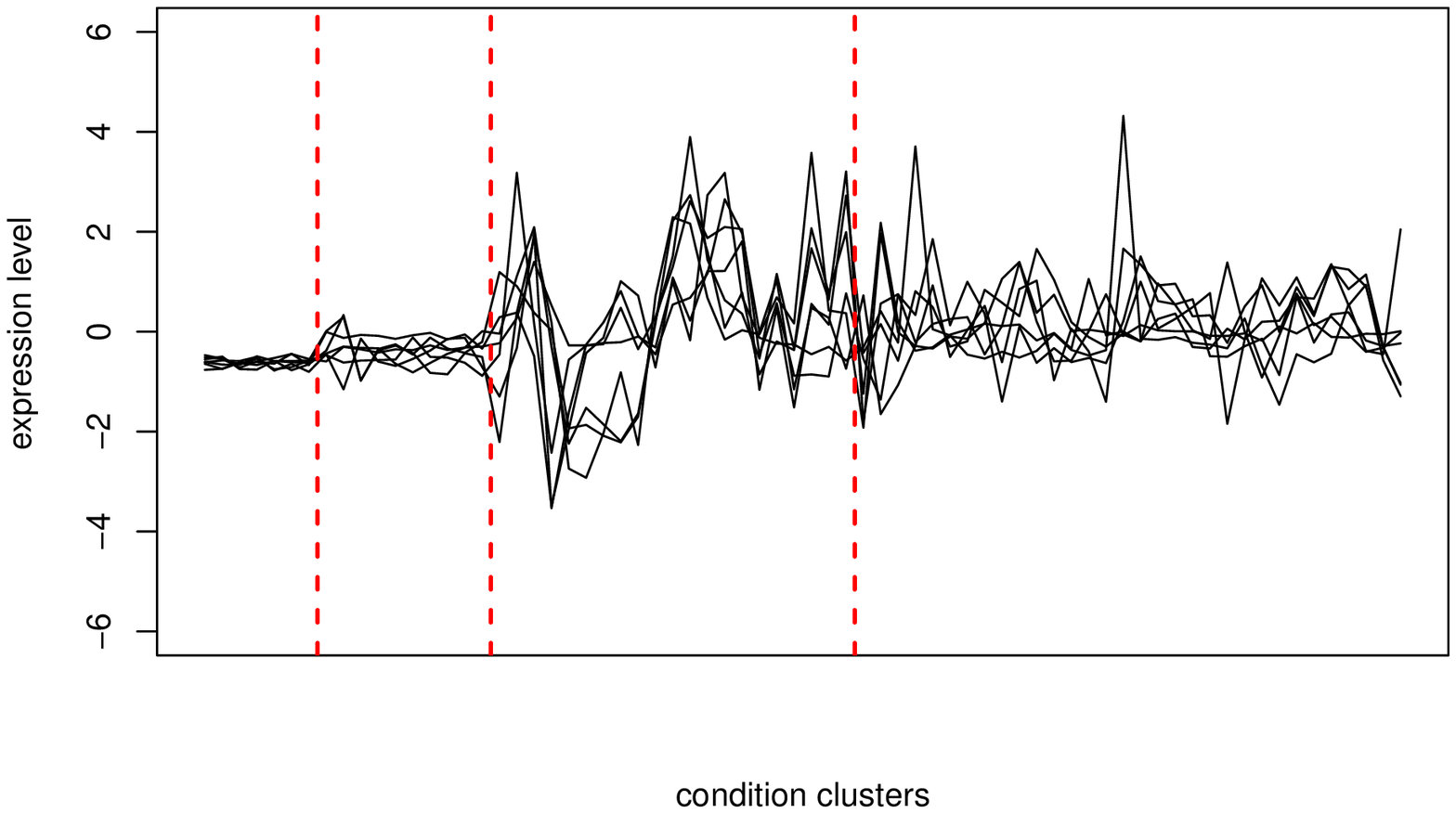}& \includegraphics[width=30mm,height=20mm]{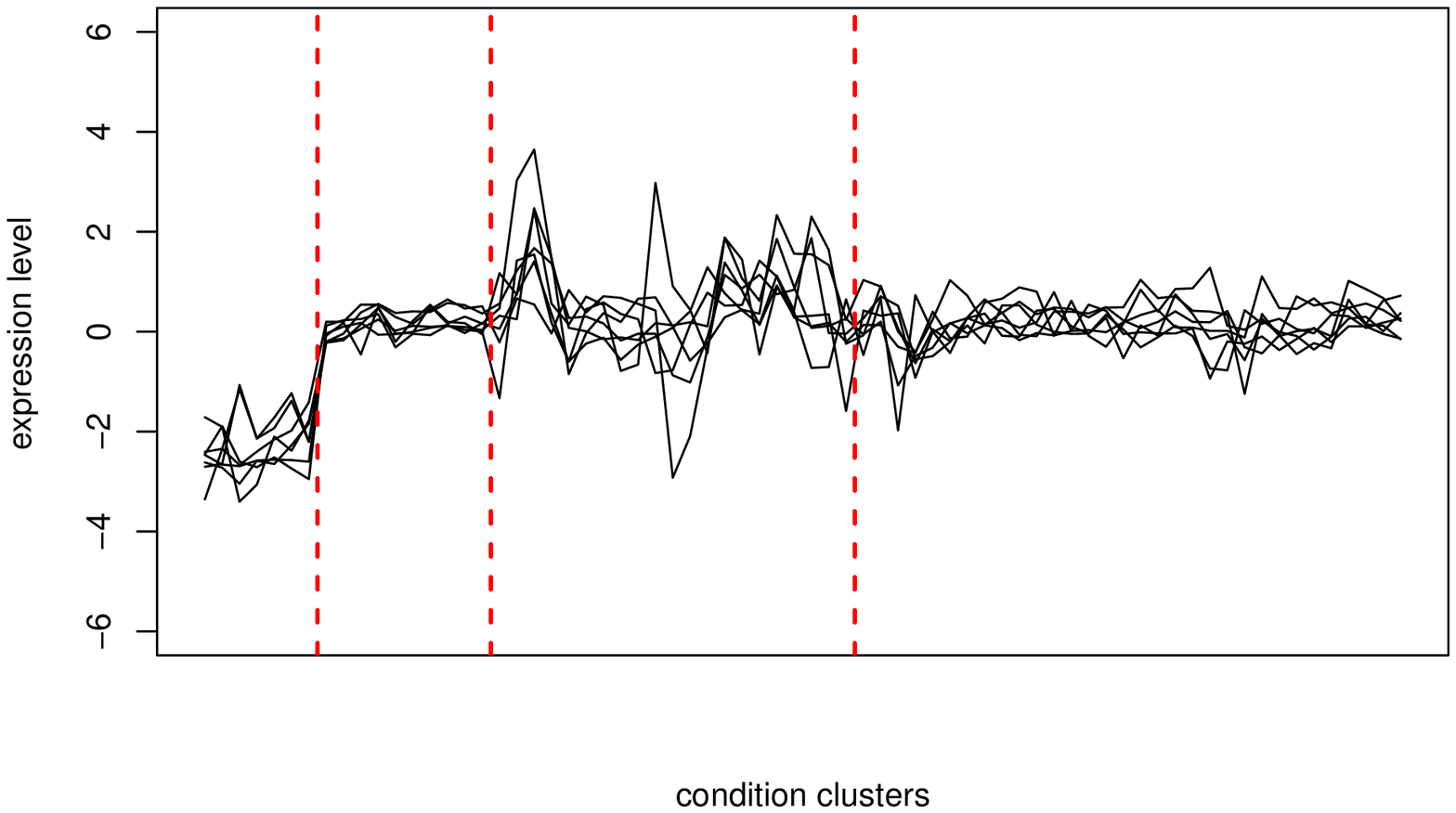}\\ \includegraphics[width=30mm,height=20mm]{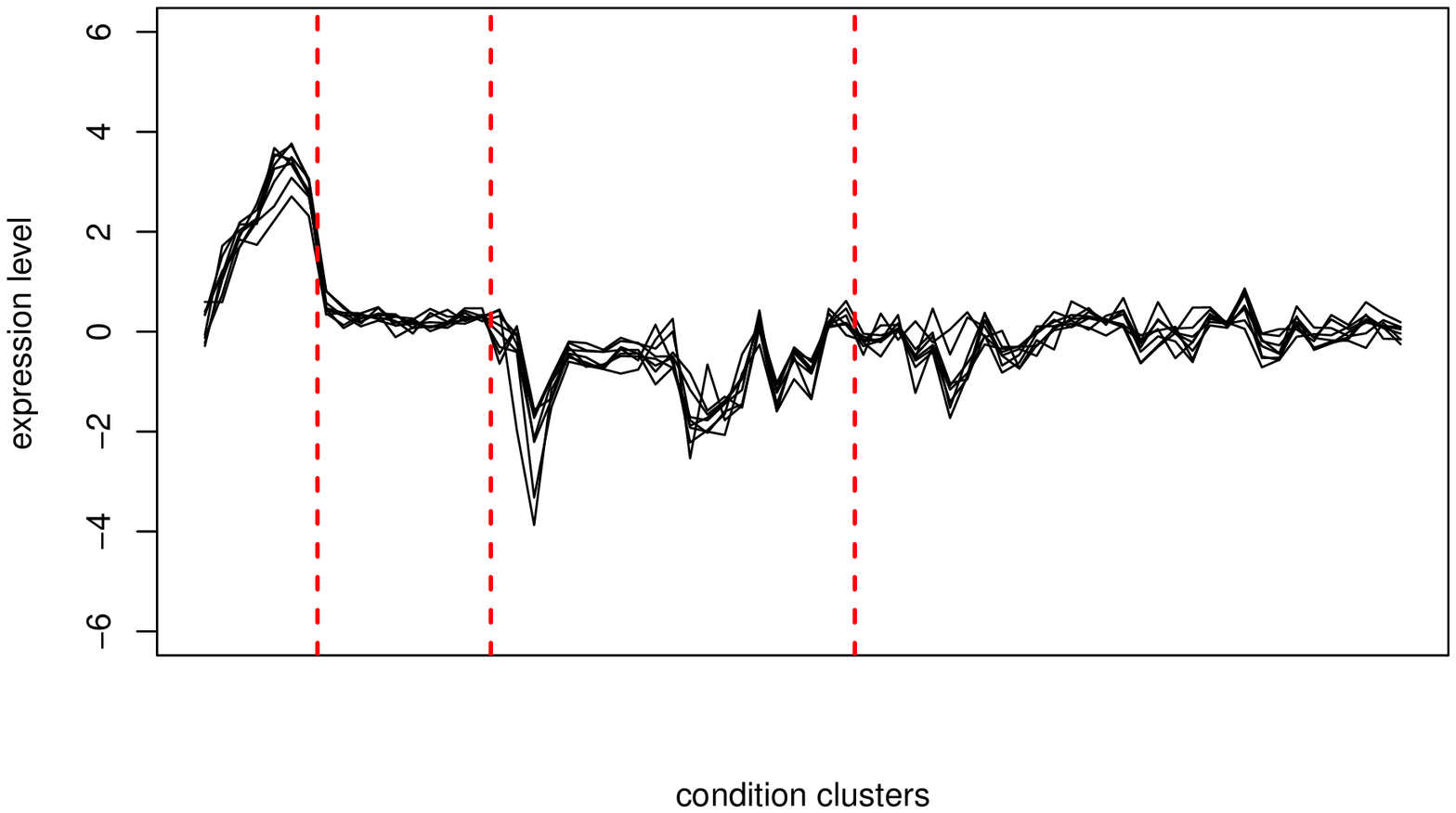} &
\includegraphics[width=30mm,height=20mm]{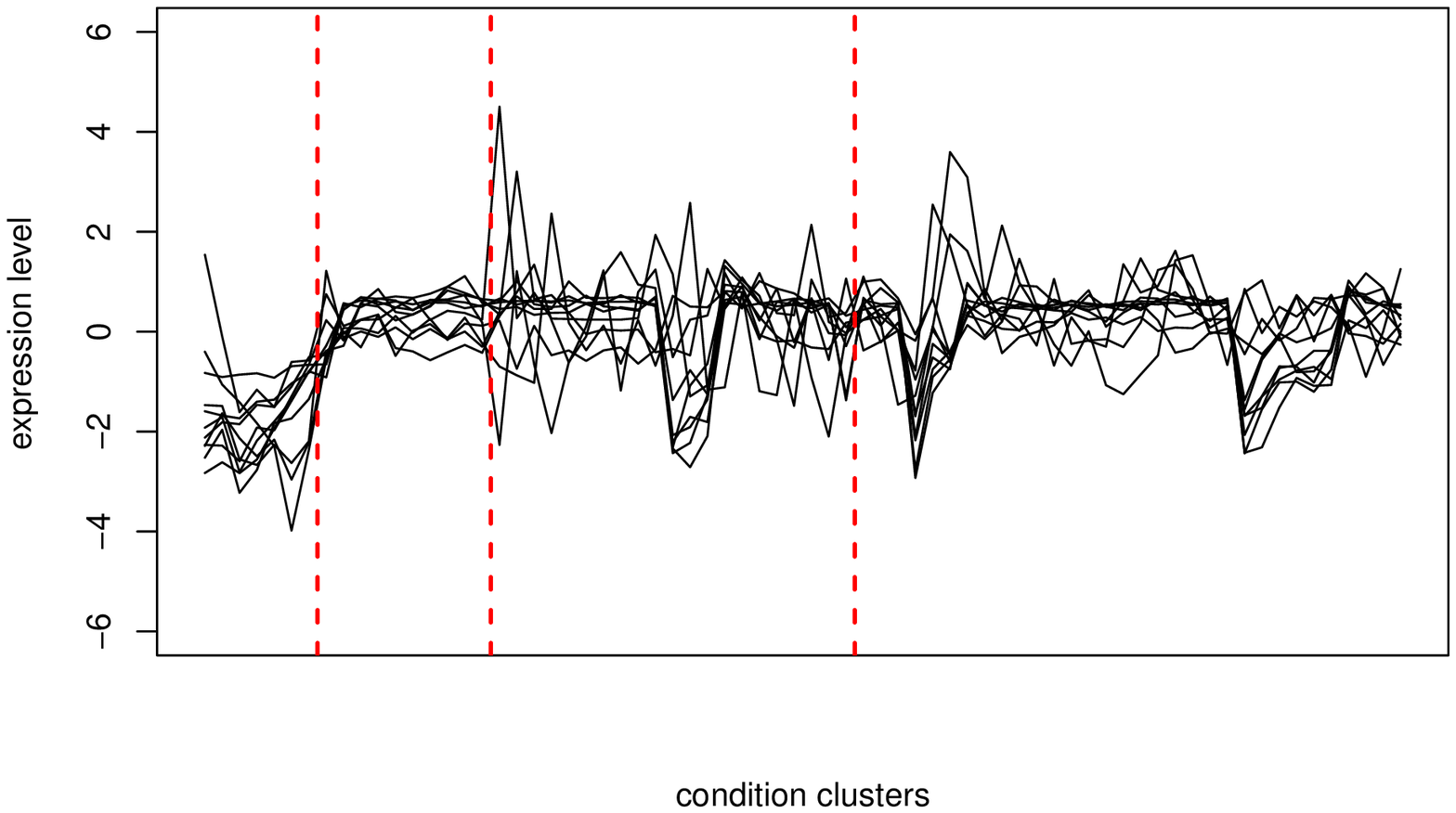} & \includegraphics[width=30mm,height=20mm]{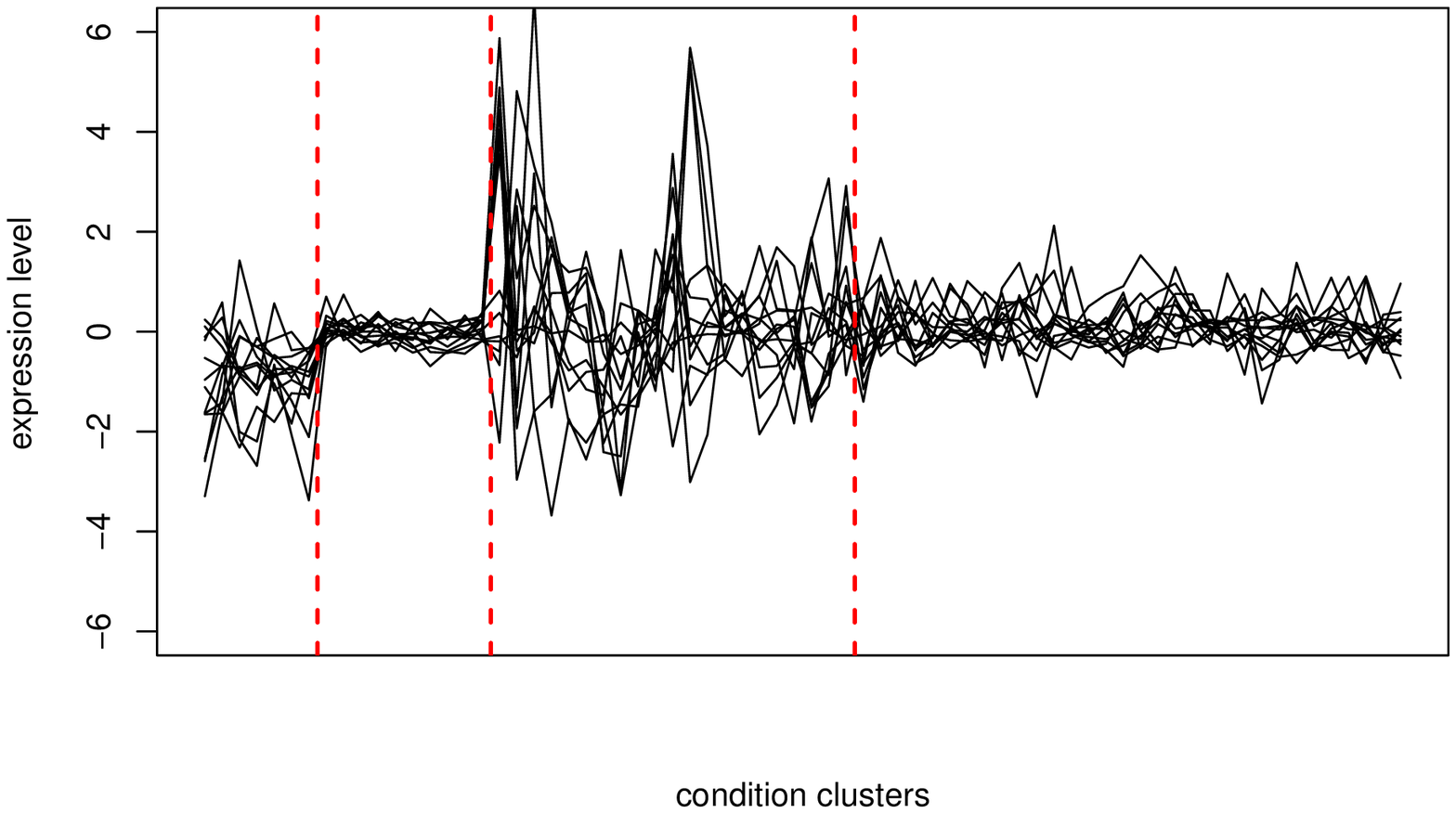} \\ \includegraphics[width=30mm,height=20mm]{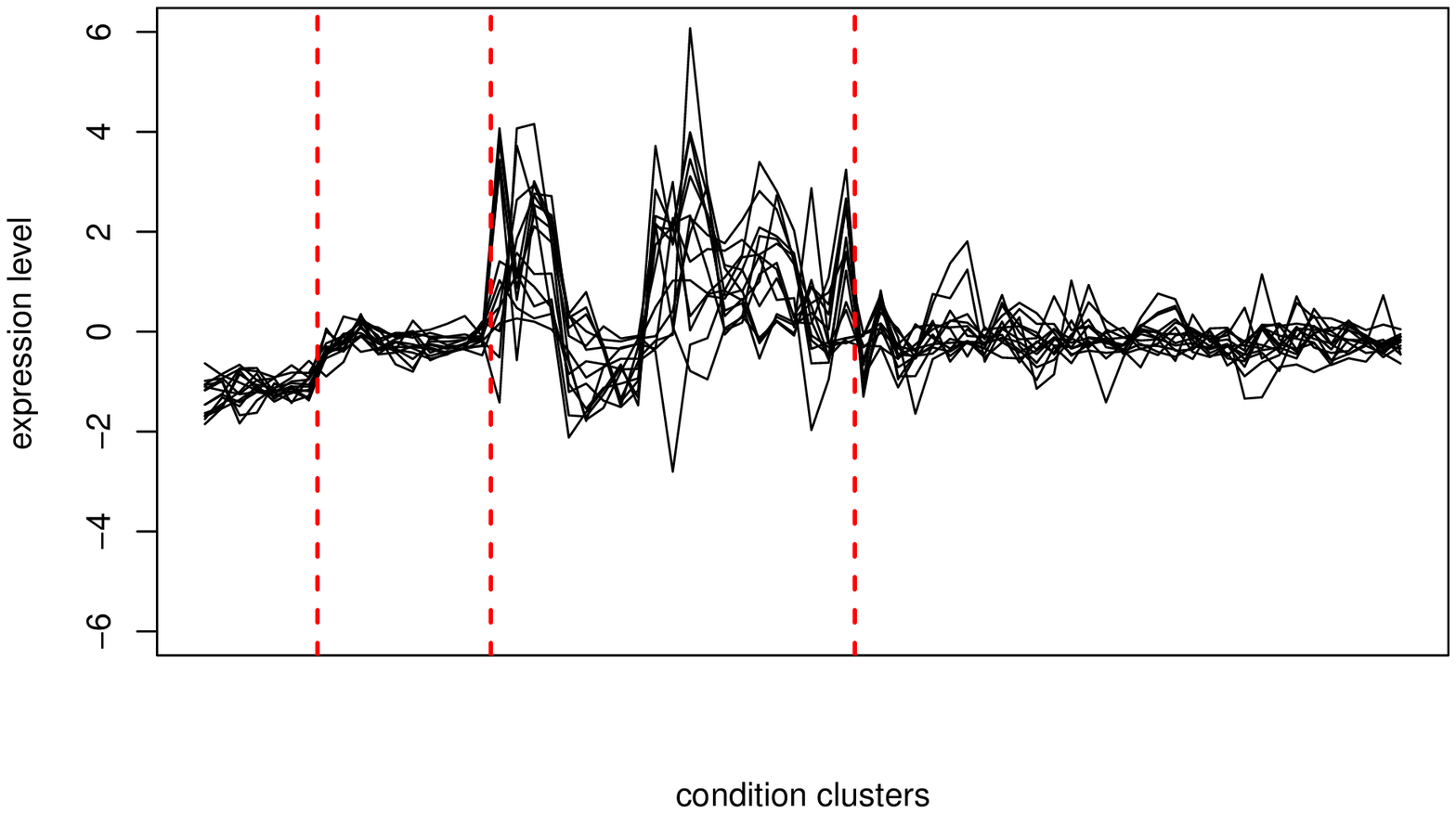}& \includegraphics[width=30mm,height=20mm]{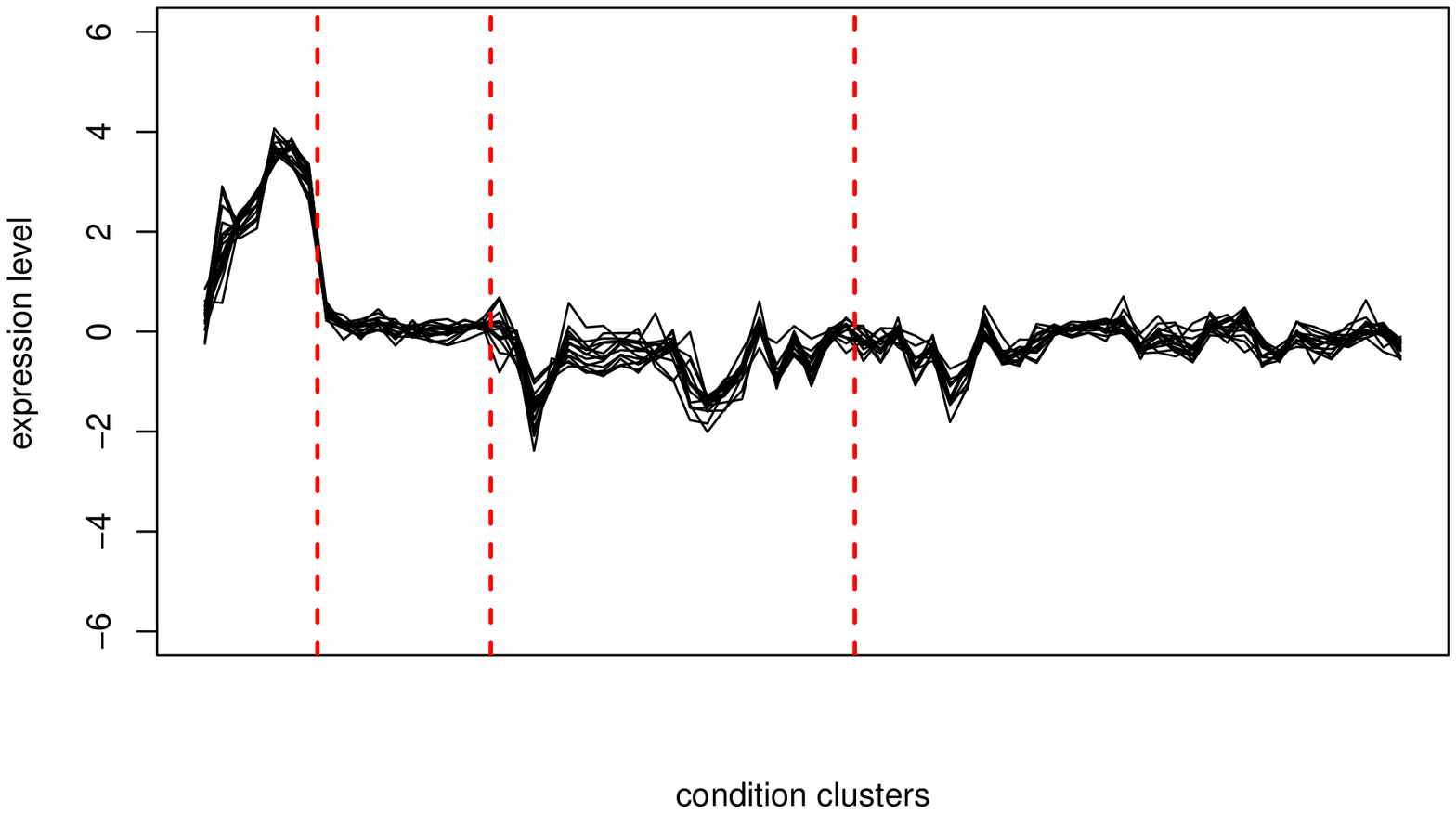} &
\includegraphics[width=30mm,height=20mm]{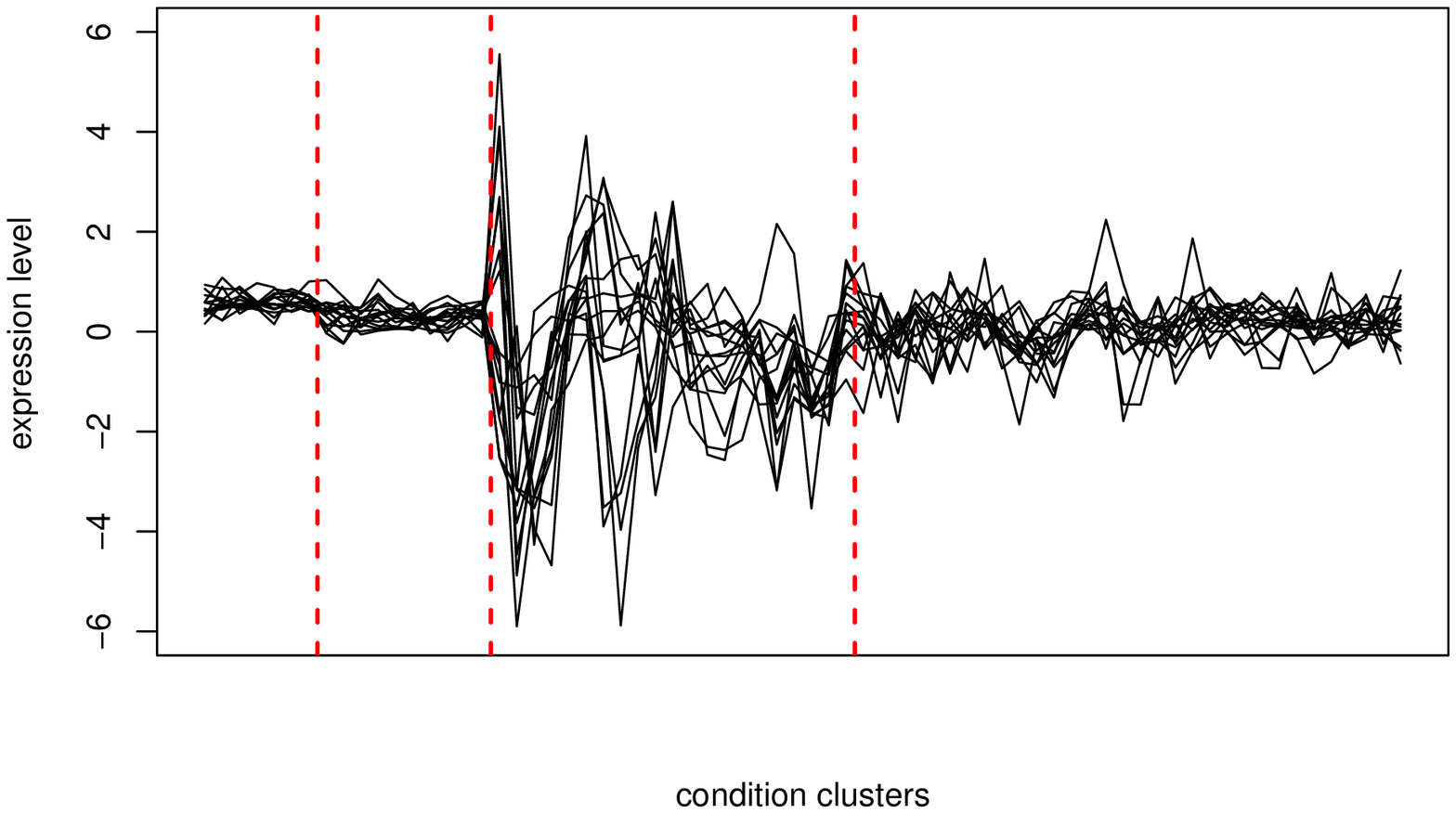} \\ \includegraphics[width=30mm,height=20mm]{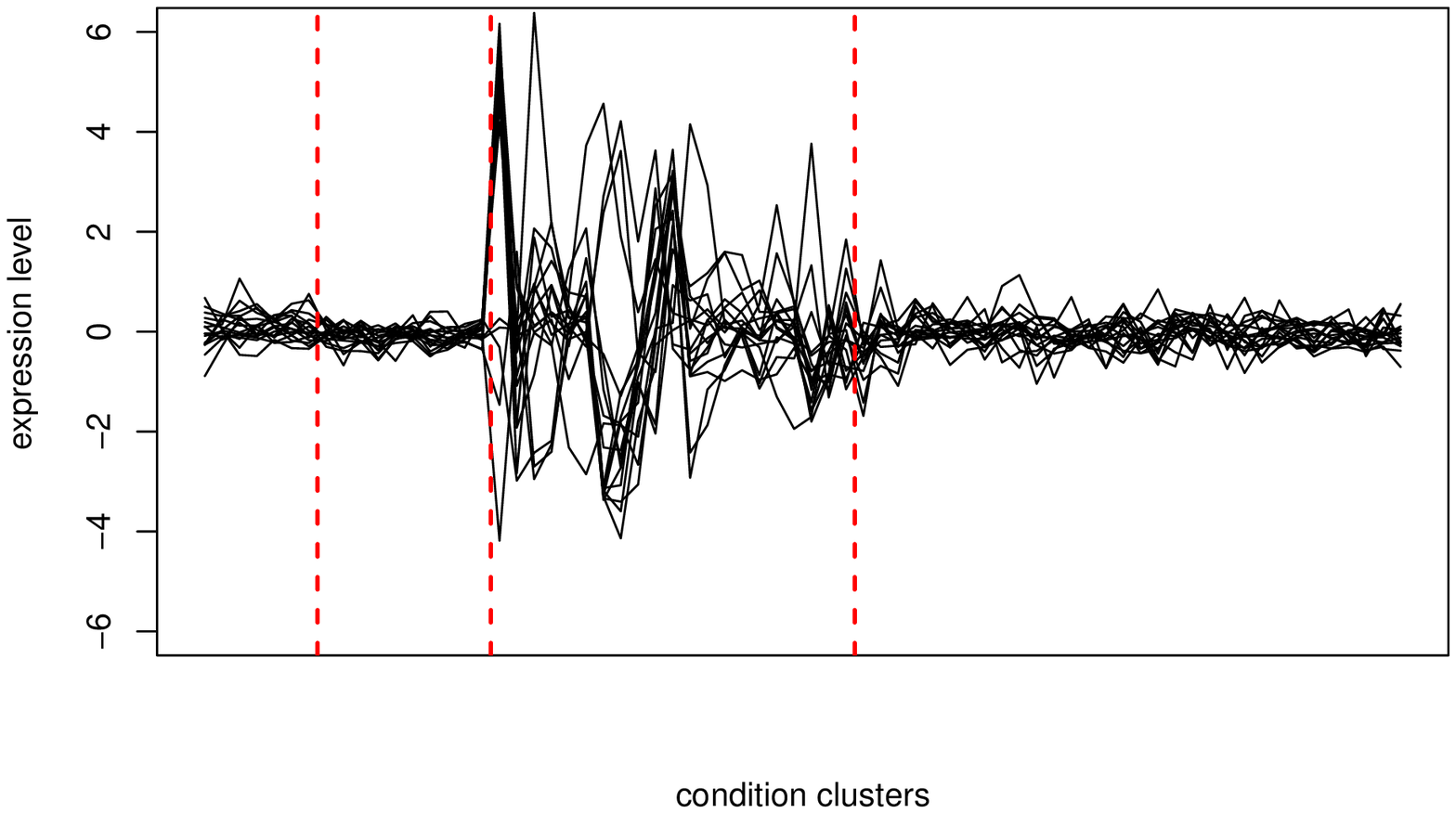}& \includegraphics[width=30mm,height=20mm]{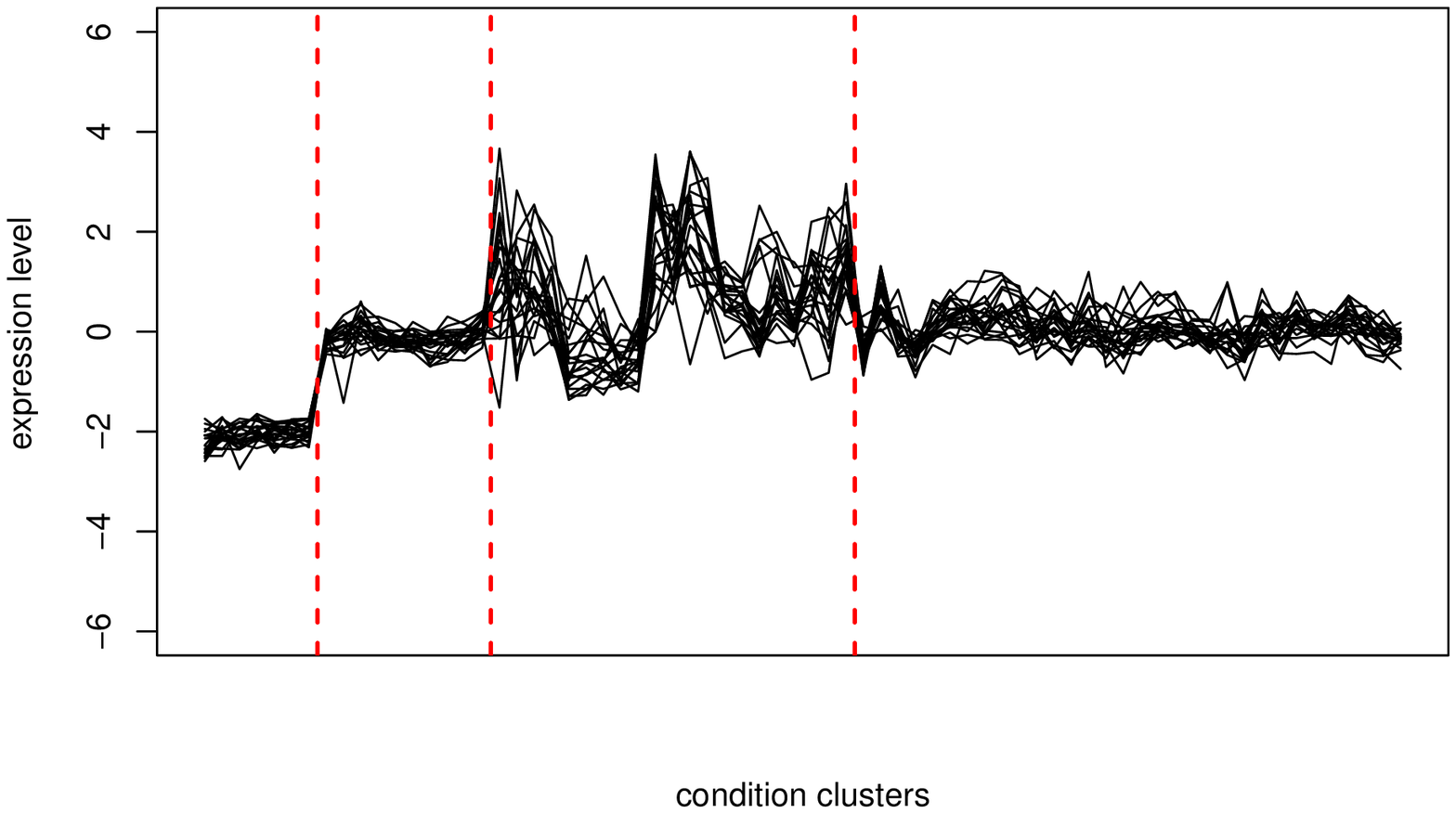}& \includegraphics[width=30mm,height=20mm]{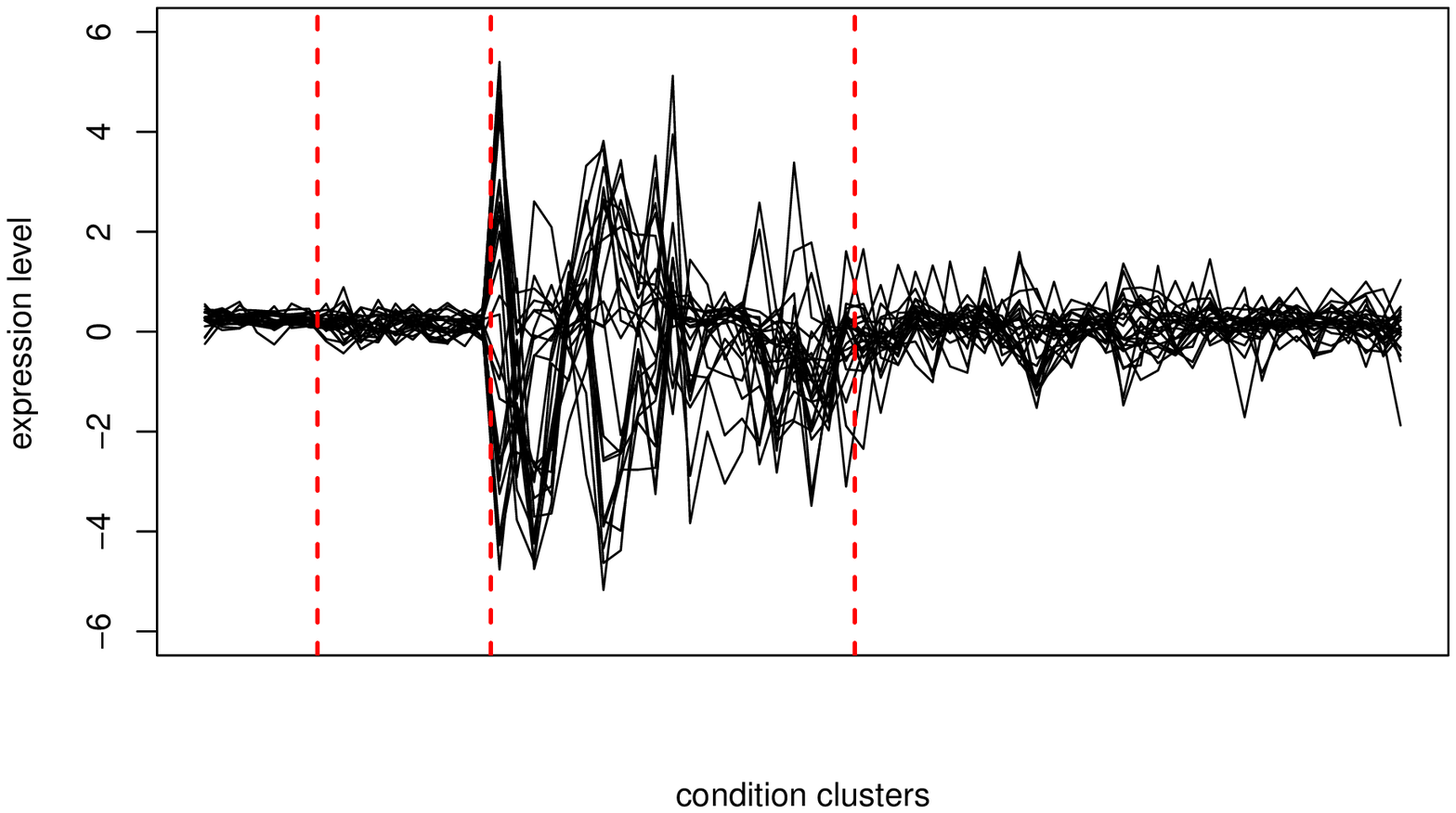} \\
\includegraphics[width=30mm,height=20mm]{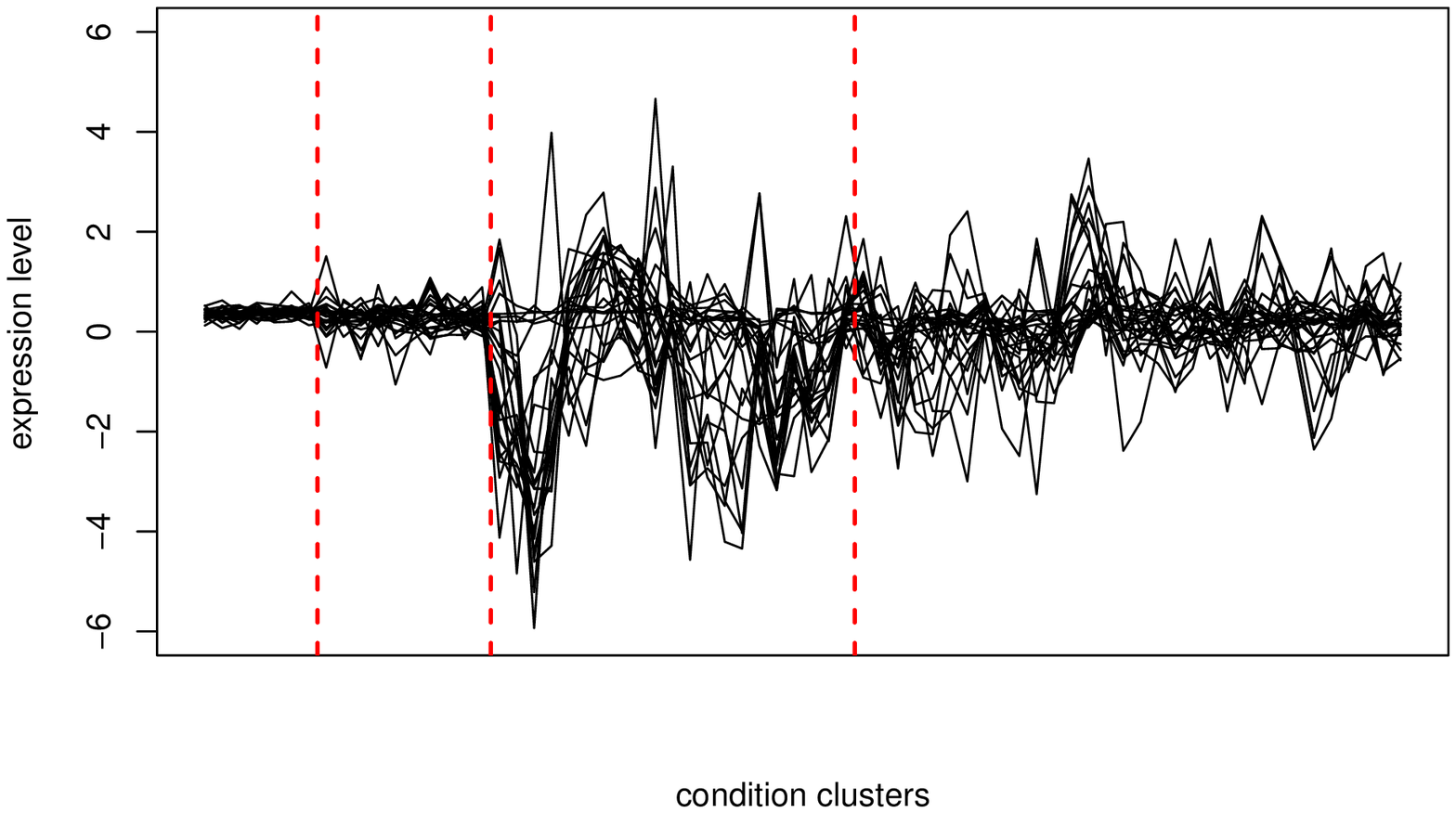} & \includegraphics[width=30mm,height=20mm]{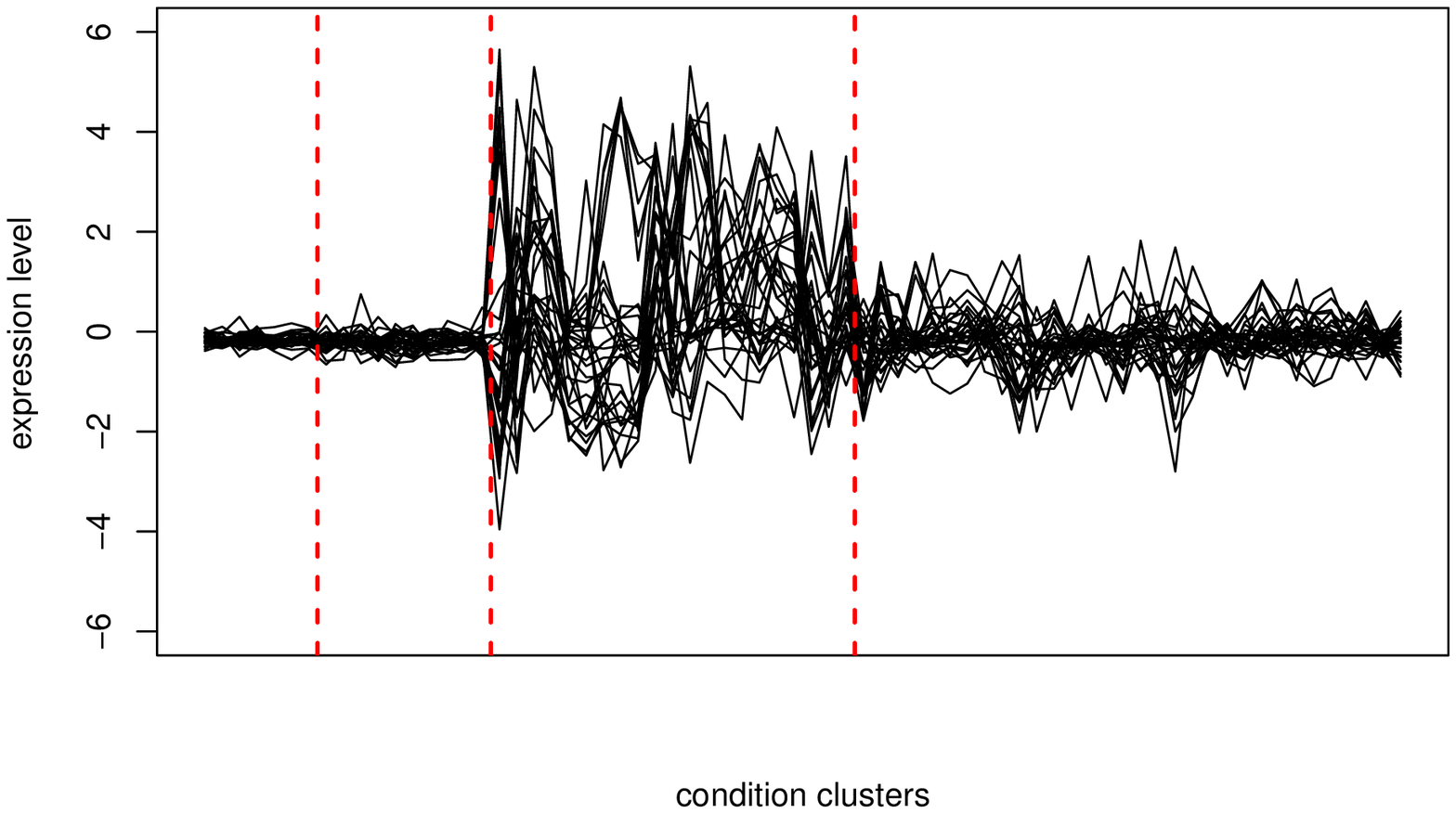}& \includegraphics[width=30mm,height=20mm]{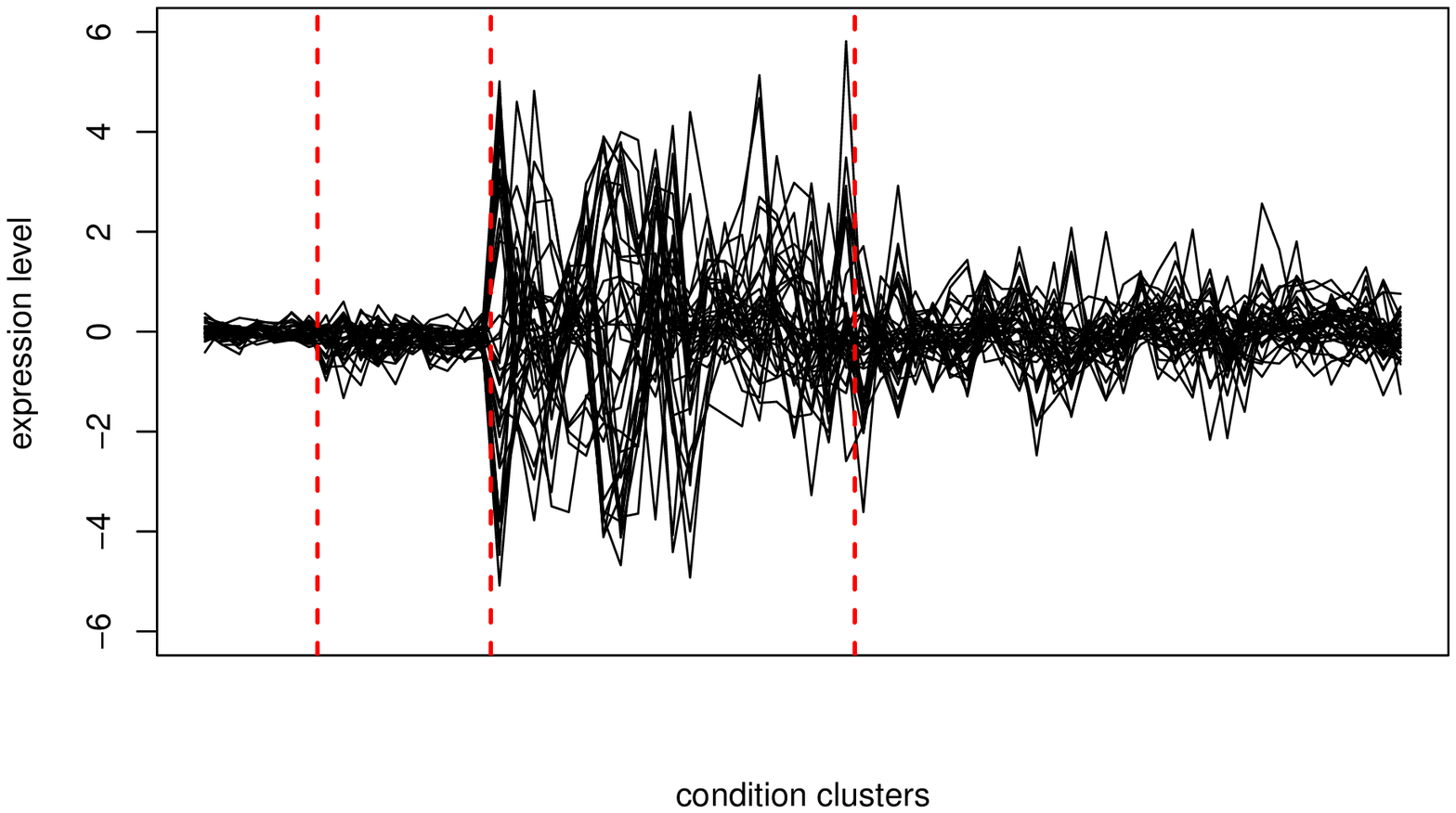}
\end{array}$
\end{center}
\caption{Selection of row clusters from the MAP clustering. Each plot corresponds to a different row cluster. Each profile (black line) gives the gene expression level for over all conditions. Conditions are arranged by condition cluster membership. The condition clusters are separated by the red dashed line.}\label{fig:bicat_clust}
\end{figure}

\section{Conclusion}

We have considered a collapsed Bayesian extension of the Latent Block Model of~\citeasnoun{Govaert08}. We showed how an MCMC sampler could be used to sample both the cluster model and the cluster memberships when clustering a data matrix into blocks. The approach was demonstrated on simulated data and two real data examples. The application to simulated data suggested that the sampler's performance deteriorates as clusters become less distinguishable. We applied the sampler to Congressional voting records from the U.S. senate. It was shown to perform well in isolating clusters of congressmen and ``yay'', ``nay'' votes in the data. In the second real data example, we used the sampler for analysis of a DNA microarray experiment. This demonstrated that there can be considerable uncertainty in the number of clusters in certain situations. Knowing even the range of possible models may be difficult. Results from the microarray experiment demonstrated that clusters could be found by a search strategy with a probabilistic basis using the collapsed LBM sampler. Overall, the approach seems to be a robust way to block cluster a data matrix. The user need only specify prior hyperparameters and priors on the number of clusters. Code implementing the sampler written in the C language is available at \texttt{www.ucd.ie/statdept/jwyse}. This code can be easily modified to experiment with different priors on the number of clusters.

\section*{Acknowledgements}

The authors would like to thank Brendan Murphy for helpful advice on aspects of this paper. Nial Friel's research was supported by a Science Foundation Ireland Research Frontiers Program grant, 09/RFP/MTH2199.

\section*{Appendix}
\subsection*{Appendix A: Calculation of the posterior} \label{sec:Acalc}

Writing out all posterior terms longhand, assuming $\boldomega \sim \mbox{Dirichlet}(\alpha,\dots,\alpha)$ and $\boldrho\sim\mbox{Dirichlet}(\beta,\dots,\beta)$ {\it a priori} gives
\begin{eqnarray*}
\pi(K,G,\bz,\bw,\boldomega,\boldrho,\Theta|Y) &\propto& \pi(K) \pi(G) \frac{\Gamma\{\alpha K\}}{\Gamma\{\alpha\}^K }\frac{\Gamma\{\beta G\}}{\Gamma\{\beta\}^G}\prod_{k=1}^K \omega_k^{n_k} \, \prod_{g=1}^G \rho_g^{m_g} \, \prod_{k=1}^K \prod_{g=1}^G \prod_{i:z_i=k} \prod_{j:w_j=g} p(y_{ij}|\theta_{kg}) \\
 & & \qquad \times \prod_{k=1}^K \prod_{g=1}^G \pi(\theta_{kg}) \, \prod_{k=1}^K \omega_k^{\alpha-1} \,\prod_{g=1}^G \rho_g^{\beta-1} \\
 & = & \pi(K) \pi(G)\frac{\Gamma\{\alpha K\}}{\Gamma\{\alpha\}^K }\frac{\Gamma\{\beta G\}}{\Gamma\{\beta\}^G} \prod_{k=1}^K \omega_k^{n_k+\alpha-1} \, \prod_{g=1}^G \rho_g^{m_g+\beta-1} \\
 & & \qquad \times \prod_{k=1}^K \prod_{g=1}^G \pi(\theta_{kg}) \prod_{i:z_i=k} \prod_{j:w_j=g} p(y_{ij}|\theta_{kg})
\end{eqnarray*}
Integrating the left and right hand sides of the above with respect to $\boldomega,\boldrho$ and $\Theta$ gives
\[
\pi(K,G,\bz,\bw|Y) \propto \pi(K) \pi(G) \frac{\Gamma\{\alpha K\}\prod_{k=1}^K \Gamma\{n_k+\alpha\}}{\Gamma\{\alpha\}^K \Gamma\{n+\alpha K\}}\frac{\Gamma\{\beta G\}\prod_{g=1}^G \Gamma\{m_g+\beta\}}{\Gamma\{\beta\}^G \Gamma\{m+\beta G\}}\prod_{k=1}^K \prod_{g=1}^G  M_{kg}
\]
where
\[
 M_{kg} = \int \pi(\theta_{kg}) \prod_{i:z_i=k} \prod_{j:w_j=g} p(y_{ij}|\theta_{kg}) \, \mbox{d}\theta_{kg}
\]

\subsection*{Appendix B: Calculation of $M_{kg}$}\label{sec:Amargs}

\subsubsection*{Bernoulli model for binary data}

Assume that $\Pr(y_{ij}=1|z_i=k,w_j=g)=\theta_{kg}$. We take a $\mbox{Beta}(\gamma,\delta)$ prior on $\theta_{kg}$. Then
\begin{eqnarray*}
\pi(\theta_{kg}) \prod_{i:z_i=k} \prod_{j:w_j=g} p(y_{ij}|\theta_{kg}) & = &\frac{\Gamma\{\gamma+\delta\}}{\Gamma\{\gamma\}\Gamma\{\delta\}} \theta_{kg}^{\gamma-1}\left(1-\theta_{kg}\right)^{\delta-1}\prod_{i:z_i=k} \prod_{j:w_j=g} \theta_{kg}^{y_{ij}}\left(1-\theta_{kg}\right)^{1-y_{ij}}\\
& =& \frac{\Gamma\{\gamma+\delta\}}{\Gamma\{\gamma\}\Gamma\{\delta\}} \theta_{kg}^{s_{kg}+\gamma-1}\left(1-\theta_{kg}\right)^{n_k m_g - s_{kg}+\delta-1}
\end{eqnarray*}
Integrating the left and right hand side of this with respect to $\theta_{kg}$ gives
\[
M_{kg} = \frac{\Gamma\{\gamma+\delta\}}{\Gamma\{\gamma\}\Gamma\{\delta\}} \frac{\Gamma\left\{ s_{kg}+\gamma\right\}\Gamma\left\{n_k m_g - s_{kg}+\delta\right\}}{\Gamma\{n_k m_g + \gamma+\delta\}}
\]

\subsubsection*{Gaussian model for continuous data}

Assume $y_{ij}|z_i=k,w_j=g \sim \mbox{N}(\mu_{kg},\sigma_{kg}^2)$. Take the priors $\mu_{kg}  \sim  \mbox{N}(\xi,\tau^2\sigma_{kg}^2)$ and $\sigma_{kg}^2  \sim  \mbox{IG}(\delta/2,\gamma/2)$ where $\mbox{IG}(a,b)$ is the Inverse-Gamma distribution: $p(x) = \frac{b^a}{\Gamma\left(a\right)}x^{-(a+1)}\exp\{-b/x\}$. Then
\begin{eqnarray*}
&& \pi(\mu_{kg}) \pi(\sigma_{kg})\prod_{i:z_i=k} \prod_{j:w_j=g} p(y_{ij}|\mu_{kg},\sigma_{kg})  =  \frac{(\gamma/2)^{\delta/2}}{\Gamma\{\delta/2\}}\sigma_{kg}^{-2(\delta/2+1)} \exp\{-\gamma/2\sigma_{kg}^2\} \\ & & \times (2\pi \tau^2 \sigma_{kg}^2)^{-1/2} \exp\{-(\mu_{kg}-\xi)^2/2\tau^2\sigma_{kg}^2\} \\ & &
\times (2 \pi \sigma_{kg}^2)^{-n_k m_g /2} \exp\left\{-(ss_{kg}-2\mu_{kg} s_{kg} + n_k m_g \mu_{kg}^2)/2\sigma_{kg}^2 \right\}
\end{eqnarray*}
Completing the square on $\mu_{kg}$ and integrating with respect to it gives
\begin{eqnarray*}
&& (2\pi)^{-n_k m_g/2}\frac{(\gamma/2)^{\delta/2}}{\Gamma\{\delta/2\}}\sigma^{-2\left((n_k m_g + \delta)/2+1 \right)}\\ &&
\times  (n_k m_g\tau^2 + 1)^{-1/2}\exp \left\{-\frac{1}{2\sigma_{kg}^2}\left(ss_{kg} -\frac{\tau^2(s_{kg}+\xi/\tau^2)^2}{n_k m_g\tau^2 + 1} + \frac{\xi^2}{\tau^2} + \gamma\right)\right\}.
\end{eqnarray*}
Finally, integrating with respect to $\sigma_{kg}^2$ and tidying up gives
\begin{eqnarray*}
M_{kg} = \frac{\gamma^{\delta/2}\,\Gamma\{(n_k m_g + \delta)/2\}}{\pi^{n_k m_g/2}\Gamma\{\delta/2\}\left(n_k m_g \tau^2 +1\right)^{1/2}}\left(ss_{kg} - \frac{\tau^2 \left(s_{kg} + \xi/\tau^2 \right)^2}{n_k m_g \tau^2 + 1}+ \frac{\xi^2}{\tau^2}+\gamma\right)^{-(n_k m_g + \delta)/2}.
\end{eqnarray*}

\subsection*{Appendix C: Cost matrix for undoing label swiching} \label{appendix_label_switching}

The cost matrix for processing the vector $\bz^{(T)}$ is
\[
C(k_1,k_2) = \sum_{t=1}^{T-1} \sum_{i=1}^n \mbox{I}\left\{z_i^{(t)} \ne k_1, z_i^{(T)} = k_2 \right\}.
\]
The more $\bz^{(T)}$ disagrees with the vectors already processed, the higher this cost will be. This is made clearer by rewriting the general entry of the cost matrix:
\begin{eqnarray}
C(k_1,k_2) & = & \sum_{t=1}^{T-1} \sum_{i=1}^n \left(1-\mbox{I}\left\{z_i^{(t)} = k_1, z_i^{(T)} = k_2 \right\}\right) \nonumber\\
						& = & n(T-1) - \sum_{t=1}^{T-1} \sum_{i=1}^n \mbox{I}\left\{z_i^{(t)} = k_1, z_i^{(T)} = k_2 \right\} \nonumber\\
						& = & n(T-1) - \sum_{i=1}^n N_i(T-1,k_1)\mbox{I}\{z_i^{(T)}=k_2\} \label{eq:rewritten_cost}
\end{eqnarray}
where $N_i(T-1,k_1)$ gives the number of processed samples up to $z^{(T-1)}$ which have given label $k_1$ to row $i$. For the sake of discussion, consider processing a sample where no label switching has occurred, $K$ is fixed and there are no changes in labels from one MCMC sample to the next. In this case the costs will be
\begin{eqnarray*}
C(k,k) &=& n(T-1) - \sum_{i=1}^n N_i(T-1,k)\mbox{I}\{z_i^{(T)}=k\} \\
       &= &  n(T-1)- n_k(T-1) \\
       & = & (n-n_k)(T-1)
\end{eqnarray*}
and for $k'\ne k$
\begin{eqnarray*}
C(k,k') &=& n(T-1) - \sum_{i=1}^n N_i(T-1,k)\mbox{I}\{z_i^{(T)}=k'\} \\
       &= &  n(T-1)- 0 \\
       & = & n(T-1).
\end{eqnarray*}
A cost of 0 could only be obtained when all rows have the same label, that is, when there is no clustering. Of course this discussion simplifies the problem somewhat. The key is in finding a permutation of the labels to minimize all costs. This permutation is found by solving the square assignment problem using the algorithm of~\citeasnoun{Carpaneto80} in our case. Finally, we note that (\ref{eq:rewritten_cost}) can be exploited to give an online post processing procedure. Define the $K\times n$ matrix $S^{(T-1)}$ with general entry $S_{ki}^{(T-1)} = \sum_{t=1}^{T-1} \mbox{I}\{z_i^{(t)}=k\}$. Then we have $C(k,k')= n(T-1) - \sum_{i=1}^n  S_{k i}^{(T-1)} I(z_i^{(T)}=k')$. After calling the square assignment algorithm and permuting the labels $\bz^{(T)}$ according to its solution, we can update $S$, using $S_{ki}^{(T)} = S_{ki}^{(T-1)}+  \mbox{I}\{z_i^{(T)} = k\}$.	

\bibliography{bibliography2}

\end{document}